\documentclass[a4paper,fleqn]{cas-sc}

\usepackage[numbers,sort,compress]{natbib}

\begin{document}

\title[mode = title]{Spectral Decomposition of Liquid Viscosity into Instantaneous Normal Modes}
\shorttitle{Spectral Decomposition of Liquid Viscosity}
\shortauthors{Long-Zhou Huang et~al.}

\author[1,2]{Long-Zhou Huang}
\credit{Methodology, Software, Visualization, Data curation, Writing - original draft}

\author[3]{Bingyu Cui}
\credit{Methodology, Software, Visualization, Data curation, Writing - original draft}

\author[1,2]{Min-Qiang Jiang}
\ead{mqjiang@imech.ac.cn}
\cormark[1]
\cortext[1]{Corresponding author}
\credit{Conceptualization, Supervision, Writing - review \& editing, Funding acquisition}

\author[4]{Matteo Baggioli}
\ead{b.matteo@sjtu.edu.cn}
\cormark[2]
\cortext[2]{Corresponding author}
\credit{Conceptualization, Supervision, Writing - review \& editing, Funding acquisition}

\author[1,2]{Yun-Jiang Wang}
\ead{yjwang@imech.ac.cn}
\cormark[3]
\cortext[3]{Corresponding author}
\credit{Conceptualization, Supervision, Writing - review \& editing, Funding acquisition}

\address[1]{State Key Laboratory of Nonlinear Mechanics, Institute of Mechanics, Chinese Academy of Sciences, Beijing 100190, China.}
\address[2]{School of Engineering Science, University of Chinese Academy of Sciences, Beijing 100049, China.}
\address[3]{School of Science and Engineering, The Chinese University of Hong Kong, Shenzhen, Guangdong 518172, PR China.}
\address[4]{Wilczek Quantum Center, School of Physics and Astronomy, Shanghai Jiao Tong University, Shanghai 200240, China \& Shanghai Research Center for Quantum Sciences, Shanghai 201315, China.}

\begin{abstract}
Viscosity, the resistance of a liquid to flow, is driven by atomic-scale friction but its microscopic origin remains poorly understood. We use a theoretical framework based on nonaffine linear response to decompose the viscosity of metallic and model liquids into contributions from individual instantaneous normal modes (INMs). Our approach reveals excellent agreement with simulations and exposes the specific excitations that govern viscous dynamics. Above the mode-coupling temperature ($T_{\text{MC}}$), viscosity is controlled by unstable localized INMs (ULINMs), which act as precursors for diffusive momentum transport. Below $T_{\text{MC}}$, we find a dynamical crossover where stable modes govern viscosity, a behavior consistent with a transition in the potential energy landscape from saddle-dominated to minima-dominated dynamics. We also propose a quantitative model connecting viscosity with ULINMs in both Arrhenius and non-Arrhenius regimes. This work provides a spectral decomposition of liquid viscosity, identifying the atomic modes responsible for it and opening a path to predict it from elementary excitations.
\end{abstract}

\begin{keywords}
Liquid Viscosity \sep Instantaneous Normal Modes \sep nonaffine \sep Statistical mechanics
\end{keywords}

\maketitle



\section{Introduction}

Viscosity is a fundamental transport coefficient in fluid dynamics \cite{viswanath2007viscosity} that reflects the ability of dissipating momentum and determines the resistance to finite rate deformations \cite{landau1987fluid}. In the hydrodynamic regime, viscosity governs the flow of physical systems at very different energy scales, from honey \cite{garcia2005viscosity} to electrons in ultra-clean metallic systems \cite{10.1063/PT.3.4497} and ultra-relativistic plasma of quarks and gluons \cite{Shuryak:2003xe}. From a macroscopic point of view, shear viscosity is defined by the linear response relation between shear stress and shear strain rate, that can be formalized using Green-Kubo formalism \cite{green1954markoff,kubo1957statistical}. This method allows for a direct estimate of the viscosity based on the time correlation function of shear stress, but its application is often hindered by the inevitable presence of hydrodynamic long-time tails \cite{PhysRevLett.18.988}. Another established method to compute the viscosity is based on the Einstein-Stokes relation between the self-diffusion constant of a probe particle and the mobility of the fluid in which the latter is flowing \cite{Stokes_2009,Broadbent1928InvestigationsOT} (see also \cite{10.1063/5.0188081} for recent developments). However, it is well-known that, as temperature approaches the glass transition, this relation breaks down hindering the applicability of this method.

Moreover, it is important to stress that both the Green-Kubo formula and the Einstein-Stokes relationship do not fully elucidate the microscopic physical mechanisms underlying momentum transport in terms of individual atomic motion. In particular, what is the direct link between microscopic dynamics and macroscopic viscosity in a fluid remains a topic of intense investigation. In fact, this issue is part of a bigger problem that is the lack of an atomic-scale microscopic description of liquid dynamics and thermodynamics \cite{trachenko2023theory,moon2024heat}, which is mostly due to the absence of a well-defined normal mode formalism for liquids, as achieved on the contrary for crystalline solids \cite{PhysRevE.108.014601}. In this context of liquids, the existence of normal modes was already proposed by Zwanzig in 1967 \cite{PhysRev.156.190}, consistent with Maxwell's intuition \cite{maxwell1867iv} that liquids present solid characteristics at short time scales. These ideas were later formalized within the so-called instantaneous normal mode (INM) approach \cite{keyes1997instantaneous,keyes-2005}, that is based on the diagonalization of the Hessian matrix in instantaneous liquid configurations. Because of the absence of a well-defined equilibrium configuration in liquids, this procedure inevitably leads to the appearance of negative eigenvalues \cite{rahman1976molecular}, or equivalently purely imaginary frequencies, that correspond to regions of the potential energy landscape with negative local curvature. Modes corresponding to positive eigenvalues are stable, while those with purely imaginary frequencies are called unstable INMs. The physical meaning of these unstable INMs remains matter of debate, contributing to the famous question of ``\textit{what INMs are and are not}'' \cite{stratt1995instantaneous}. 

Partially answering this question, in a series of works \cite{keyes1994unstable,doi:10.1063/1.469947}, Keyes proposed that unstable INMs might provide the fundamental blocks to achieve a microscopic derivation of the liquid self-diffusion constant, highlighting their physical nature as relaxational processes mediated by jumps over potential barriers. It was later clarified that only unstable delocalized INMs contribute to the self-diffusion constant \cite{PhysRevLett.74.936,10.1063/1.3701564} (see also \cite{10.1063/1.474822,PhysRevLett.78.2385,PhysRevLett.84.4605,PhysRevE.64.036102,PhysRevE.65.026125,10.1063/1.474822,10.1063/1.474968} for related works and different definitions of this ``diffusive'' subset of INMs), providing a first connection between microscopic excitations and macroscopic transport in liquids, but leaving the physical meaning of the localized unstable modes unclear.

In this work, we investigate whether a microscopic definition for another fundamental transport coefficient in liquids -- shear viscosity-- can be achieved using a normal mode approach. Combining extensive numerical simulations with a recent theoretical approach based on nonaffine linear viscoelastic response, we propose a spectral decomposition of liquid viscosity into instantaneous normal modes. In doing so, we will reveal the so-far unknown physical significance of unstable localized modes and their essential role for the microscopic origin of liquid viscosity.

\section{Methodology} 

\subsection{Theoretical model}

According to the theoretical formalism presented in Ref.~\cite{PhysRevE.108.044101}, that is based on the prior formulation of nonaffine viscoelastic response \cite{lemaitre2006sum}, the shear viscosity derived from the imaginary part of the complex shear modulus is given by
\begin{equation}
\eta=3 \rho \,\tilde{\upsilon}(0) \int \frac{g\left(\omega\right)
\Gamma\left(\omega\right)}{m^{2} \omega^{4}}\, d \omega
\label{eq01},
\end{equation}
where $\rho =N/V$ represents the number density, $N$ is the number of atoms and $V$ is the volume of the system. Additionally, $g(\omega)$ is the density of states (DOS); $m$ is the atomic mass; $\Gamma(\omega)$ is the affine force field correlator \cite{lemaitre2006sum} (refer to \textit{Appendix} \ref{afaf}  for more details); $\tilde{\upsilon} (0)$ is the zero-frequency limit of the spectral density, $\tilde{\upsilon}(\omega)=\int_{0}^{\infty}\upsilon(t)e^{-i\omega t}dt$, with $\upsilon(t)$ the memory kernel arising from the coupling between tagged particles and thermal environment \cite{Cui2018,Cui2017}. Eq.~\eqref{eq01} alludes to the ability of achieving a mode decomposition of liquid viscosity, similar to what done in solids for many other thermodynamic and transport properties such as heat capacity, thermal conductivity and so on. However, particularly in the context of liquid systems, Eq.~\eqref{eq01} and its associated theoretical framework leave several key steps unspecified for accomplishing this task.

First, the definition of the DOS $g(\omega)$ in liquids is not straightforward. In our work, we propose to identify $g(\omega)$ with the INM density of states, allowing a direct connection to normal modes (that is not possible using the velocity auto-correlation function). Second, following this choice, it is necessary to define the integration range on $\omega$, that now takes values along both the real and imaginary axes. In the following, to determine the appropriate integration range, we adopt a pragmatic approach: we explore different choices and compare them directly with simulation data. Notably, previous works (e.g., \cite{10.1063/5.0272171}) have included all instantaneous normal modes, both stable and unstable, in Eq.~\eqref{eq01}. As we show below, explicit comparisons with simulation data for three different systems reveal that this choice produces results that deviate by several orders of magnitude from the actual data and it is therefore incorrect.  

Finally, to avoid introducing free parameters or phenomenologically fixed constants, we derive the memory kernel $\upsilon(t)$ from first principles, employing the projection operator formalism \cite{10.1063/1.4868653,PhysRevLett.116.147804,Jung2017IterativeRO},
\begin{equation}
\upsilon(t)=\frac{\left\langle \mathbf{F}\left(e^{-i(1-\mathcal{P})\mathcal{L}t}\mathbf{F}\right)\right\rangle}{k_\text{B}T}=\frac{\langle \mathbf{FF}_r(t)\rangle}{k_\text{B}T}=\frac{\bar{C}_{\mathbf{FF}}(t)}{k_\text{B}T}
\label{eq02},
\end{equation}
where $\mathbf{F}$ is the true force acting on the tagged particle. The random force $\mathbf{F}_{r}=e^{-i(1-\mathcal{P})\mathcal{L}t}\mathbf{F}$, $-i\mathcal{L}$ is the Liouvillian operator corresponding to the unperturbed dynamics and $\mathcal{P}$ is the Mori projection operator along the direction of the velocity. Consequently, the zero-frequency limit of spectral density can be simply expressed as
\begin{equation}
\tilde{\upsilon}(0)=\int\limits_0^\infty\frac{\bar{C}_{\mathbf{FF}}(t)}{k_\text{B}T}dt
\label{eq03}.
\end{equation}
For more technical details, we refer to \textit{Appendix} \ref{propro}.

All in all, once the range of integration on $\omega$ in Eq.~\eqref{eq01} is fixed, our procedure will not involve any free fitting parameter.

\subsection{Molecular dynamics simulation}

In order to provide an as much as possible comprehensive analysis, in this study, three simulation systems are considered: (I) a $\mathrm{Cu_{50}Zr_{50}}$ metallic liquid with $T_\mathrm{g}= 690$ K \cite{10.1063/1.5131500}, (II) a covalently bonded Fe$_{80}$P$_{20}$ metallic liquid with $T_\mathrm{g}= 805$ K \cite{Ackland_2004}\color{black}, and (III) a Kob-Andersen (KA) simulation model with $T_\mathrm{g}= 0.39$ \cite{PhysRevE.51.4626,PhysRevE.52.4134}.

Two potential potential  functions are respectively used to simulate the interactions in a $\mathrm{Cu_{50}Zr_{50}}$ \cite{10.1063/1.5131500} and $\mathrm{Fe_{80}P_{20}}$ \cite{Ackland_2004} metallic glass with a simulation box of 4000 atoms. The liquid and glass samples are prepared with the standard heating-cooling method. First, a crystalline $\mathrm{Cu_{50}Zr_{50}}$ and $\mathrm{Fe_{80}P_{20}}$ phase are respectively heated from 300 K to 2300 K and 500 K to 2500 K for melting. And the liquid is thermally equilibrated for 1 ns at high temperature which is much longer than the $\alpha$ relaxation time. Then the equilibrium liquid was quenched to low temperature with constant cooling rate of $10^{10}$ K/s. The above sample preparation process was carried out under the NPT ensemble with periodic boundary condition. The Nose-Hoover thermostat was used to control temperature and the barostat was used to keep pressure to zero \cite{10.1063/1.447334}. The kinetic properties were estimated using standard microcanonical ensemble after sufficient equilibration at each specific temperature. For statistical purpose, ten and five independent samples are  respectively simulated for $\mathrm{Cu_{50}Zr_{50}}$ and $\mathrm{Fe_{80}P_{20}}$ to reduce the fluctuation of physical quantities. The velocity Verlet algorithm was used to integrate Newton's equations of motion with a time step of 1 fs.

The Kob-Andersen model is binary Lennard-Jones mixture introduced by Kob and Andersen \cite{PhysRevE.51.4626,PhysRevE.52.4134} and defined by the following LJ potential,
\begin{equation}
U_{a b}(r)=4 Y_{a b}\left[\left(\frac{X_{a b}}{r}\right)^{12}-\left(\frac{X_{a b}}{r}\right)^{6}\right]
\label{eqs01},
\end{equation}
where $a,b\in \left \{ A,B \right \} $; here $Y_{AA} = 1.0, X_{AA} = 1.0, Y_{AB}= 1.5, X_{AB} = 0.8, Y_{BB} = 0.5$, and $X_{BB} = 0.88$. The KA model consists of particles A and B with the same mass $m$ in a ratio of 80:20. We use reduced units, where length is in the unit of $X_{AA}$, temperature $Y_{AA}/k_{\mathrm{B} }$, and time $(mX_{AA}^{2}/ Y_{AA})^{1/2}$. Under periodic boundary conditions, the system contains 4000 particles. Under the NVT ensemble, the fixed density is 1.185 and the cooling rate is $3.3\times 10^{-6}$. The kinetic properties were calculated using standard microcanonical ensemble. Ten independent samples were obtained to reduce the fluctuation error of the simulations. The time step is 0.005. For the KA model, the temperature $T$ will be always displayed in reduced LJ units.

\begin{figure}
\centering
\includegraphics[width=0.9\textwidth]{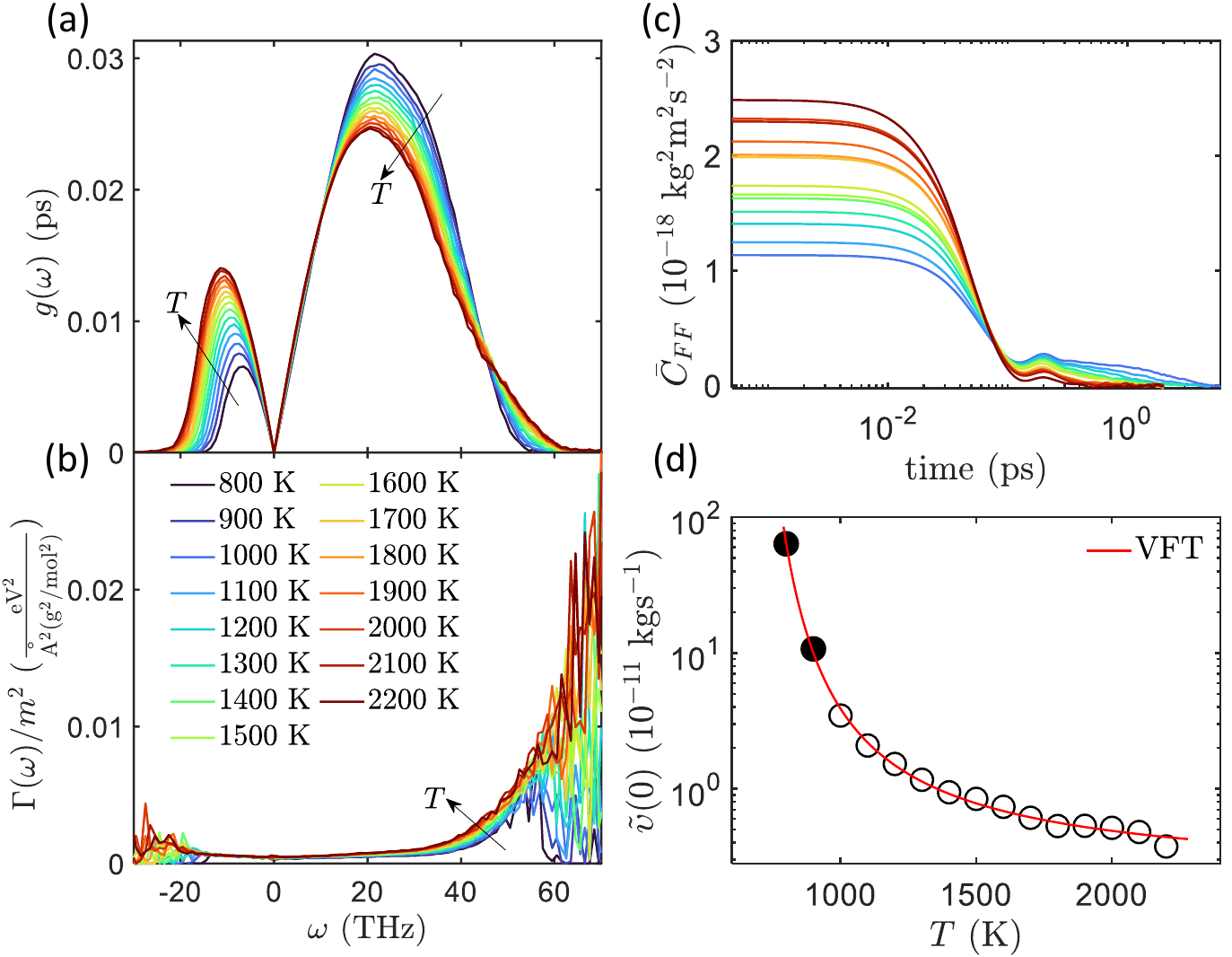}
\caption{\label{fig01}\textbf{Physical properties behind the viscosity of a Cu-Zr metallic liquid.} \textbf{(a)} INM DOS $g(\omega)$ and \textbf{(b)} affine force field correlator $\Gamma(\omega)/m^2$ versus frequency at different temperature. \textbf{(c)} Auto-correlation function of the random force $\bar{C}_{\mathbf{FF}}(t)$ versus time at different temperature. \textbf{(d)} Temperature dependence of the zero-frequency limit of spectral density $\tilde{\upsilon}(0)$. Solid black circles are calculated using Stokes-Einstein relation, while hollow circles by the projection operator technique, Eq.~\eqref{eq03}.}
\end{figure}

\section{Results and discussion}

\subsection{Normal mode description of liquid dynamics}

We performed INM analysis of the $\mathrm{Cu_{50}Zr_{50}}$ metallic liquid for temperatures between $800$ K and $2200$ K. The corresponding INM DOS is shown in Fig.~\ref{fig01}(a), following the common practice of representing imaginary frequencies along the negative axes \cite{keyes1997instantaneous}. The arrows indicate the directions along which $T$ increases. We notice that $g(\omega)$ is linear at low frequency (see \textit{Appendix} Fig. \ref{fig9}), in agreement with theoretical expectations (\textit{e.g.}, \cite{keyes1994unstable,doi:10.1073/pnas.2022303118,PhysRevE.55.6917}) and experimental observations \cite{doi:10.1021/acs.jpclett.2c00297,jin2024temperature}. 

\begin{figure}
\centering
\includegraphics[width=0.9\textwidth]{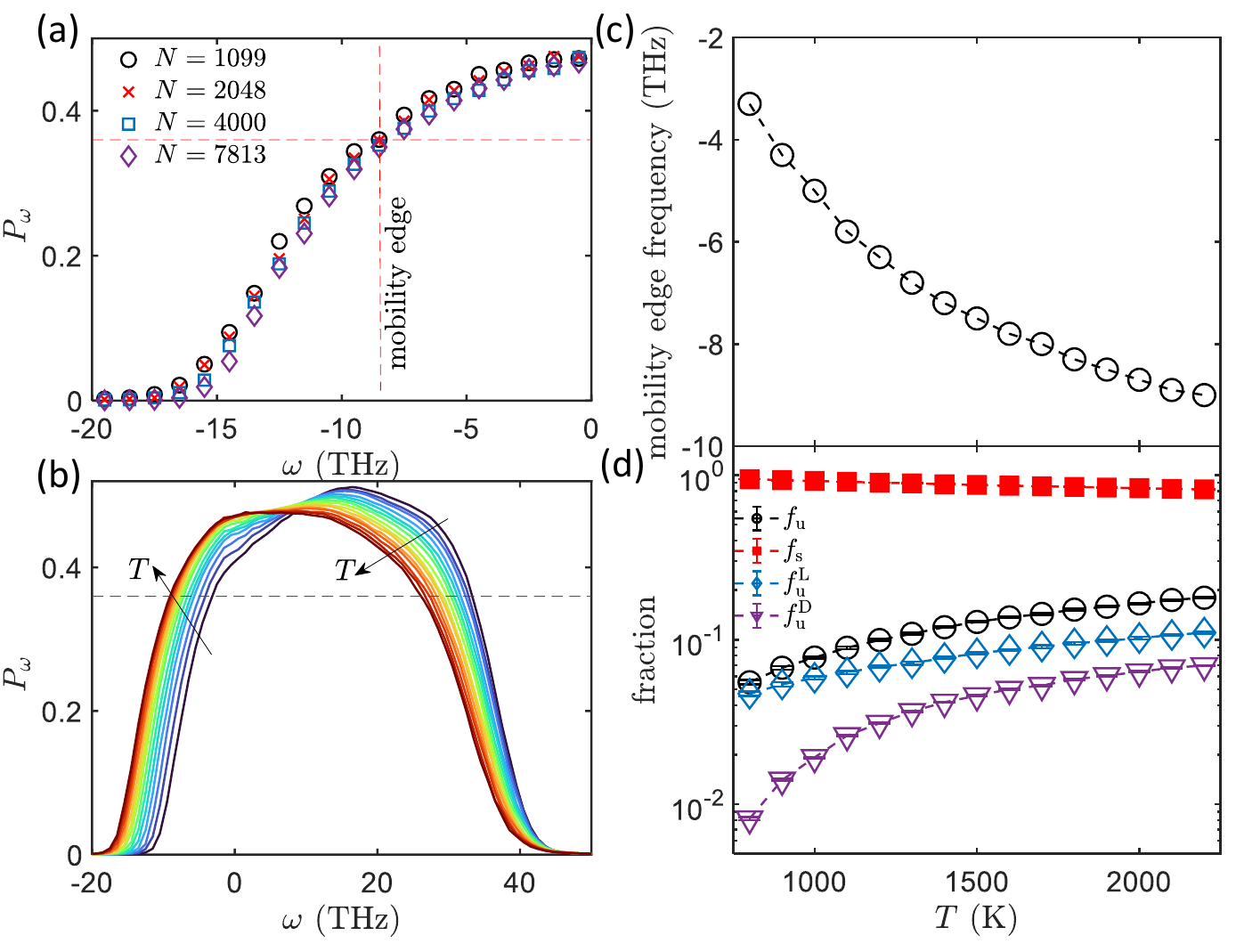}
\caption{\label{fig02}\textbf{Nature of instantaneous normal modes in a Cu-Zr metallic liquid.} \textbf{(a)} Variation of the participation ratio  as a function of (imaginary frequency) at $1800$ K upon changing the size of the simulation box. \textbf{(b)} Participation ratio versus frequency at different temperatures. The dotted horizontal line represents the mobility edge threshold $P_{\omega}=0.36$. \textbf{(c)} Temperature dependence of the mobility edge frequency (for a similar analysis on the KA model, see \textit{Appendix} \ref{prpr}). \textbf{(d)} Fraction of stable, unstable, unstable localized and unstable delocalized modes as a function of temperature.}
\end{figure}

Panels (b) and (c) of Fig.~\ref{fig01} display respectively the re-scaled affine force field versus frequency and the auto-correlation function of the random force versus time at different temperatures. We find that computing $\tilde{\upsilon}(0)$ using the Einstein's relation $k_{\mathrm{B}}T/{D}$, where $D$ is the long-range diffusion constant of the tagged particles, gives results compatible with those obtained using the projection operator technique, Eq.~\eqref{eq03}, at low-enough temperatures (see \textit{Appendix} Fig. \ref{fig13}). Hence, the Einstein's relation is used only for few data points (solid black circles in Fig.~\ref{fig01}(d)) at very low temperatures and, otherwise, $\tilde{\upsilon}(0)$ is computed using the projection operator technique. The data also follow the empirical Vogel–Fulcher–Tammann (VFT) equation \cite{vogel1921law,fulcher1925analysis,tammann1926dependence}, $A(T)=A_{0}\exp\left(\frac{B_{\mathrm{VFT}}}{T-T_{0}}\right)$, shown with a red line in the same panel. 

As anticipated, INMs can be classified into stable and unstable modes depending on the sign of the corresponding eigenvalues. The latter can be further divided into localized and delocalized, depending whether the number of particles participating in the mode is intensive or extensive. Bembenek and Laird originally argued that only unstable delocalized modes are related to diffusion \cite{PhysRevLett.74.936,10.1063/1.471147}, while unstable localized modes do not contribute to it. However, there are indications that unstable localized modes may play a fundamental role for other dynamical properties, such as structural relaxation \cite{10.1063/1.5127821,Widmer-Cooper2008}.

\begin{figure}
    \centering
    \includegraphics[width=0.9\linewidth]{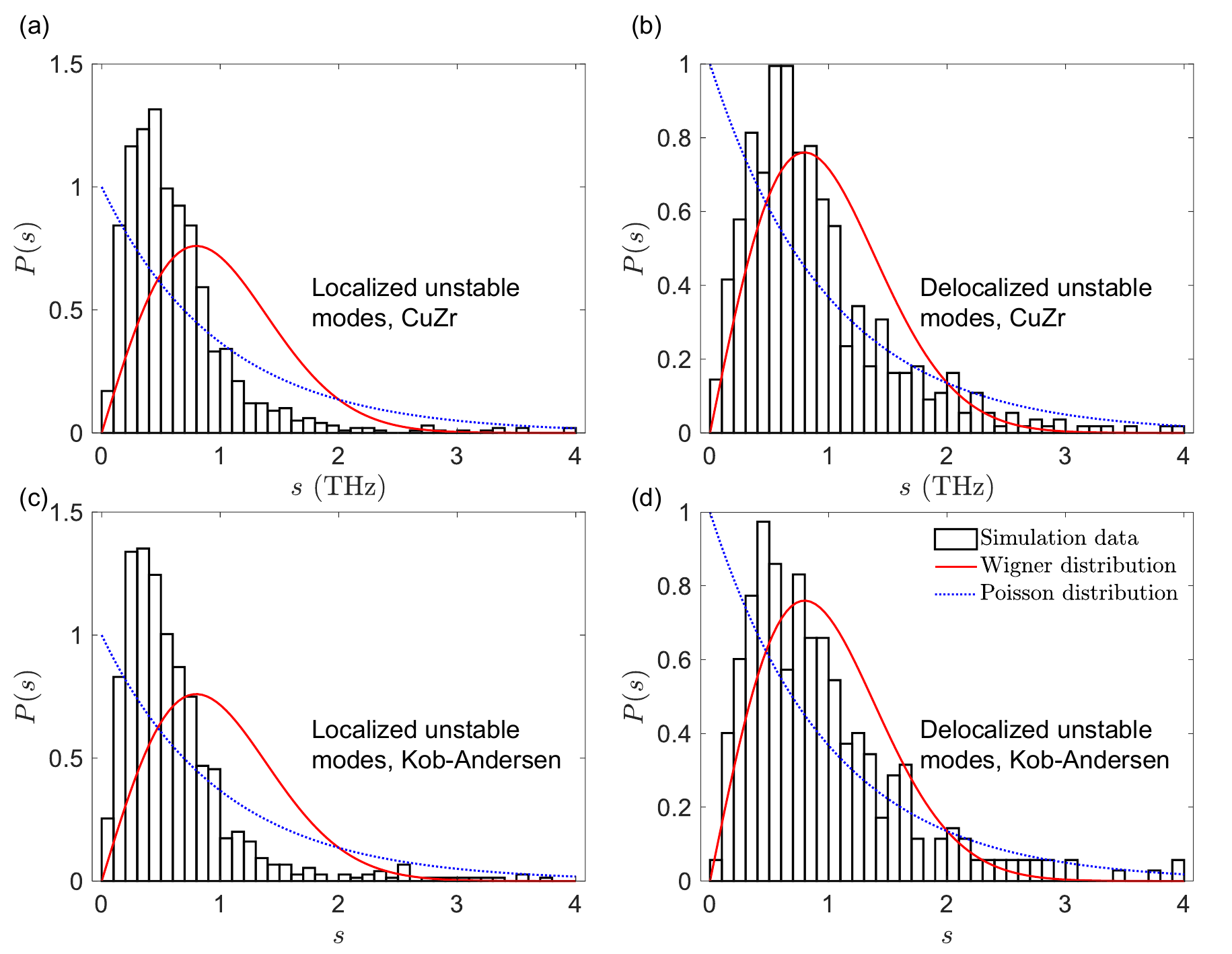}
    \caption{\textbf{Spectral statistics of unstable modes.} The distribution $P(s)$ of unfolded frequency spacings $s$. (a) Unstable localized and (b) unstable delocalized modes for CuZr liquid at $T=1500$ K. (c) Unstable localized and (d) unstable delocalized modes for Kob-Andersen liquid at $T=0.6$. The red solid and blue dashed line represent the Wigner (Eq.~\eqref{wig}) and Poisson (Eq.~\eqref{poi}) distribution, respectively.}
    \label{fig3}
\end{figure}

\begin{figure}
\centering
\includegraphics[width=0.6\textwidth]{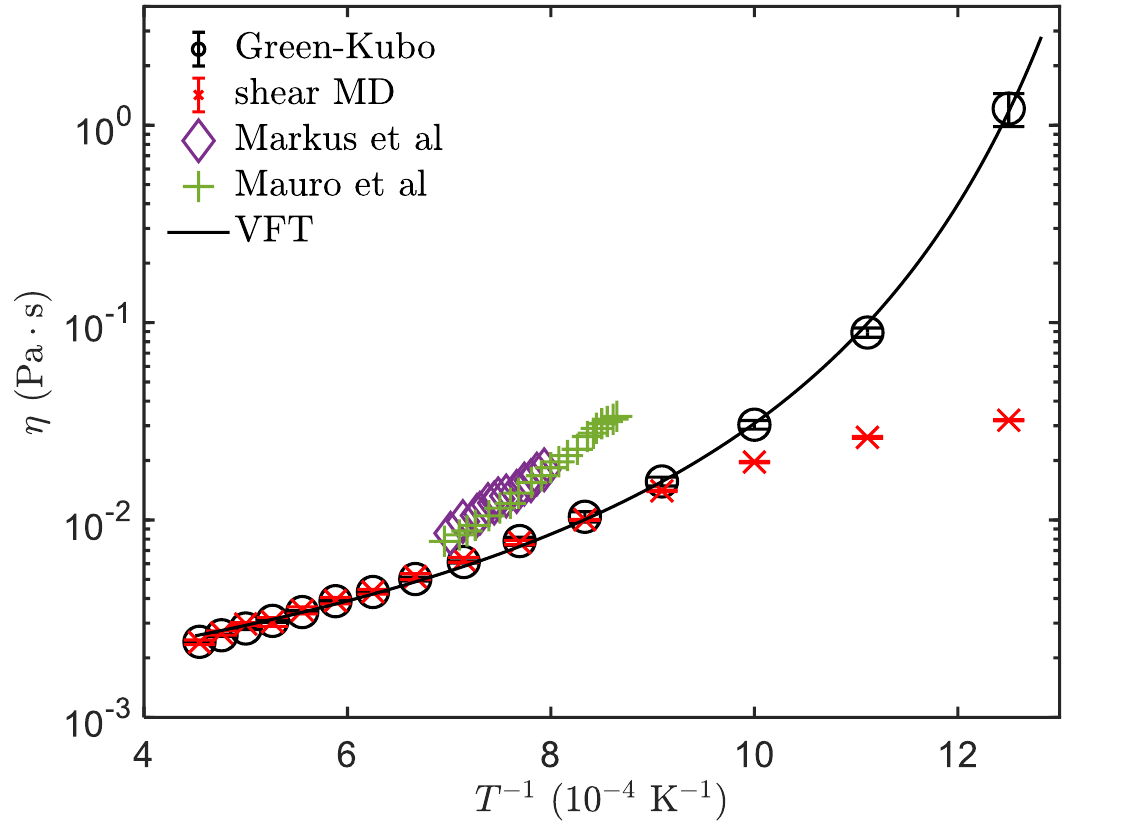}
\caption{\textbf{Viscosity of the Cu-Zr metallic liquid.} Comparison of viscosities from the Green-Kubo formula (black circles), MD non-equilibrium shear (red crosses), and experimental data from Refs. (Markus et al. \cite{mohr2019surface}) (purple diamonds) and (Mauro et al. \cite{mauro2014structural}) (green crosses). The solid line represents the VFT fit.}
\label{fig4}
\end{figure}

In order to quantify the degree of localization, we consider the participation ratio
\begin{equation}
P_{\omega}=\left [N\sum_{i=1}^{N}(\mathbf{e}_{i}^{\omega}\cdot \mathbf{e}_{i}^{\omega} )^{2}\right ]^{-1} 
\label{eq06},
\end{equation}
where $\mathbf{e}_{i}^{\omega}$ is the contribution of particle $i$ to the normalized eigenvector corresponding to eigenfrequency $\omega$. $P_{\omega}$ takes values between $1/N$ and $1$, with the former corresponding to an extremely localized mode with only one particle participating, while the latter to a fully delocalized mode involving all particles (see \textit{Appendix} Fig. \ref{fig17}) for direct visualization of these two types of modes and Ref.~\cite{mizuno2023computational} for details about $P_{\omega}$). 

In order to identify the \textit{mobility edge} separating localized and delocalized modes, we follow \cite{PhysRevLett.74.936,10.1063/1.471147} and perform a finite size scaling analysis and compute $P_{\omega}$ for different $N$. In Fig.~\ref{fig02}(a), we show $P_\omega$ as a function of frequency for the unstable branch of INMs and different system sizes. From there, we can observe that $P_\omega$ becomes system dependent below a certain threshold frequency, that for the $\mathrm{CuZr}$ glass-forming system corresponds to $P_{\omega} \approx 0.36$. Intensive modes, with $P_\omega$ below the mobility edge, are localized, while extensive modes with $P_\omega \gtrapprox 0.36$ are delocalized.

The participation ratio as a function of frequency for different temperatures is shown in Fig.~\ref{fig02}(b). By repeating the same finite size scaling analysis for different $T$, we found that the value $P_{\omega} \approx 0.36$ for the $\mathrm{Cu_{50}Zr_{50}}$ metallic liquid (while $P_{\omega} \approx 0.24$ and $P_{\omega} \approx  0.27$ for the $\mathrm{Fe_{80}P_{20}}$ metallic liquid and the KA model; see \textit{Appendix} \ref{prpr}), defining the mobility edge, is in first approximation independent of temperature (see \textit{Appendix} Fig. \ref{fig16} for a direct proof of this statement). By then searching for the corresponding (imaginary) frequency (see horizontal dashed line in Fig.~\ref{fig02}(b)), we have derived the mobility edge frequency that is plotted in Fig.~\ref{fig02}(c) as a function of temperature. We find that the mobility edge frequency increases (in absolute value) by increasing temperature, consistent with the idea that unstable modes tend to be more localized towards the glass transition temperature \cite{PhysRevLett.74.936}.

Finally, in Fig.~\ref{fig02}(d), we respectively show the fraction of stable $f_{\mathrm{s}}$, unstable $f_{\mathrm{u}}$, unstable localized $f_{\mathrm{u}}^{\mathrm{L}}$ and unstable delocalized  $f_{\mathrm{u}}^{\mathrm{D}}$ INMs in the temperature range considered. We find that the fraction of stable modes slowly decreases by increasing temperature, while all the other fractions increases with temperature following a similar trend. Once the temperature reaches the glass transition, unstable delocalized modes disappear and all unstable modes become localized, consistent with previous results \cite{10.1063/1.3701564}. 

To confirm the validity of our separation between localized and delocalized modes---which is essential for the subsequent analysis---we examine in detail the spectral statistics of each subset. Specifically, we compute the nearest-neighbor spacing distribution \( P(s) \) of the unfolded frequency spacings,
\begin{equation}
s_n = \frac{\omega_{n+1}-\omega_n}{\langle \omega_{n+1}-\omega_n \rangle}.
\end{equation}
This quantity provides a well-established diagnostic of mode localization~\cite{PhysRevE.70.061502}. Delocalized modes are expected to follow the Wigner distribution,
\begin{equation}
P(s)=\frac{\pi}{2}s\exp\!\left(-\frac{\pi}{4}s^2\right),\label{wig}
\end{equation}
whereas localized modes obey Poisson statistics,
\begin{equation}
P(s)=\exp(-s).\label{poi}
\end{equation}

The results are presented in Fig.~\ref{fig3}. For the CuZr liquid at \( T=1500\,\mathrm{K} \) and the Kob--Andersen liquid at \( T=0.6 \), the corresponding mobility-edge frequencies are \( -7.5\,\mathrm{THz} \) and \( -3.7 \), respectively. Although the agreement is not exact, Figs.~\ref{fig3}(a) and (c) show that the unstable localized modes display spacing distributions close to Poisson statistics. In contrast, Figs.~\ref{fig3}(b) and (d) indicate that the unstable delocalized modes follow distributions much closer to the Wigner form, consistent with their extended character.

Overall, this frequency-spacing analysis provides independent support for the robustness of our protocol to distinguish localized from delocalized modes.

\color{black}We have now independently derived all the physical quantities entering in Eq.~\eqref{eq01} and analyzed in detail the different types of modes over which the integration therein could be in principle performed. Before moving to the computation of the liquid viscosity, we notice that, due to the limited size of the simulation box $L$, the integration in Eq.~\eqref{eq01} will be always performed up to a lower cutoff frequency $|\omega_{\min }|=(2 \pi v_s) / L$, where the speed of sound $v_s$ is given by $\sqrt{B / \bar{\rho}}$, with $B$ the bulk modulus and $\bar{\rho}$ the mass density (see \textit{Appendix} \ref{bubu} for more details). We note that Ref.~\cite{10.1063/5.0272171} introduced a non-vanishing static shear modulus in the liquid state, whose physical basis remains unclear, resulting in a different definition of the cutoff, a choice we do not adopt here.

\begin{figure}
\includegraphics[width=\textwidth]{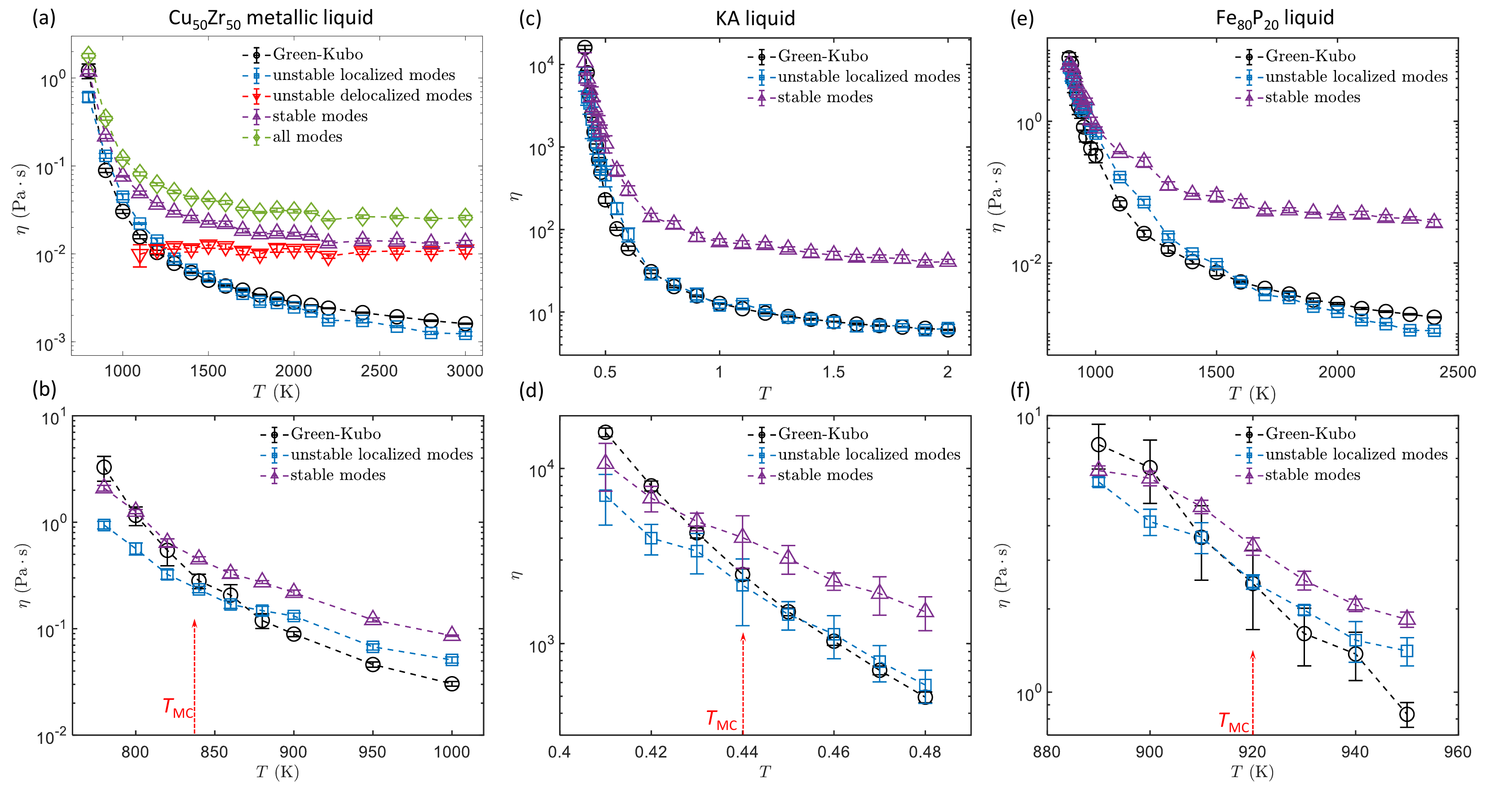}
\caption{\label{fig5}\textbf{Normal mode decomposition of liquid viscosity.} \textbf{(a)} Comparison of viscosities from Green-Kubo formula (black circles) and different INMs including unstable localized (blue symbols), unstable delocalized (red inverted triangles), stable (purple triangles) and all modes (green diamonds). \textbf{(b)} Zoom on the low-temperature regime. The red arrow locates the approximate position of the mode coupling temperature $T_{\text{MC}} \approx 840$ K. Panels \textbf{(c)-(d)} show the same analysis for the KA liquid where $T_{\text{MC}}\approx 0.44$. Finally, panels \textbf{(e)-(f)} display the same results for the covalently bonded Fe$_{80}$P$_{20}$ metallic liquid for which $T_{\text{MC}}\approx 920\ \mathrm{K}$. To avoid clutter, in panels (c) and (e) only the theoretical predictions with unstable localized modes and stable modes are presented.}
\end{figure}

\color{black}\subsection{Liquid viscosity from instantaneous normal modes}

Traditionally, the viscosity can be computed using the Green-Kubo formula \cite{green1954markoff,kubo1957statistical},
\begin{equation}
\eta=\frac{V}{k_{\mathrm{B}} T} \int\limits_{0}^{\infty} \left\langle\sigma_{\alpha \beta}(t) \sigma_{\alpha \beta}(0)\right \rangle  d t
\label{eq07},
\end{equation}
where $\sigma_{\alpha \beta}$ is the $\alpha,\beta$ component of the stress tensor (with $\alpha,\beta=x,y,x$) and $\langle \cdots \rangle$ indicates ensemble average.

Alternatively, the viscosity can be obtained from non-equilibrium molecular dynamics using the linear response relation $\eta =\sigma _{\alpha \beta}/\dot{\varepsilon}$, where $\dot{\varepsilon }$ is the shear strain rate. We will indicate this second route as \textit{shear MD}. In Fig.~\ref{fig4}, we compare the viscosity for the Cu-Zr metallic liquid calculated by different methods: black symbols corresponds to the Green-Kubo formula Eq.\eqref{eq07}, red crosses to shear MD. For comparison, we also show the experimental data from Refs. \cite{mohr2019surface}) (purple diamonds) and \cite{mauro2014structural} (green crosses). At high temperature (above $T\approx 1100$ K, $1.5$ times the glass transition temperature), Green-Kubo formula and shear MD are in good agreement. However, when temperature drops, the shear viscosity becomes strain rate dependent and the linear response formula (hence, the shear MD method) is not anymore applicable, giving unreliable predictions for $\eta$.

We are now ready to compare the viscosity obtained from simulations with the predictions of the theoretical formula Eq. \eqref{eq01}. In Fig.~\ref{fig5}(a), we show the viscosity of the Cu-Zr metallic liquid extracted via Eq. \eqref{eq07} together with the theoretical predictions from Eq. \eqref{eq01} using different sets of normal modes. As already anticipated, and incorrectly assumed in Ref.~\cite{10.1063/5.0272171}, it emerges clear that using all the instantaneous normal modes in Eq. \eqref{eq01} leads to an incorrect prediction of $\eta$ by several orders of magnitude, especially at high temperature. Most importantly, we find that the viscosity computed via the Green-Kubo formula seems in good agreement with the results of Eq.~\eqref{eq01} using only ULINMs, and both data are well described by the VFT equation. A closer inspection in the low-temperature regime, provided in Fig.~\ref{fig5}(b), reveals the emergence of a dynamical crossover between approximately $800$ K and $870$ K below which $\eta$ becomes controlled by the stable modes that correspond to regions of local positive curvature in the potential energy landscape (PEL). This crossover emerges approximately at the mode coupling temperature $T_{\text{MC}}$ of the system that is $\approx 840$ K and it is consistent with the well-established topological transition from a dynamics between basins of saddles (``\textit{saddles dominated}'') at $T>T_{\text{MC}}$ to a dynamics between basins of minima (``\textit{minima dominated}'') at $T<T_{\text{MC}}$ \cite{PhysRevLett.85.5356,PhysRevLett.85.5360}. The observed crossover in $\eta$ provides a direct manifestation of this topological transition onto the transport properties of the liquid.

To provide a more comprehensive analysis, we performed the same analysis in a KA liquid and in a covalently bonded Fe$_{80}$P$_{20}$ metallic liquid. In Fig.~\ref{fig5}(c), we present the viscosity for the KA liquid obtained from the Green-Kubo formalism and we compare it with the predictions from Eq. \eqref{eq01} using unstable localized modes and stable modes. At high temperature, $T > 0.6$ in LJ units, we find that $\eta$ is well captured by the theoretical formula using only the ULINMs. In the low temperature region, we observe a dynamical crossover below which the viscosity becomes dominated by stable modes rather than unstable localized ones. Fig.~\ref{fig5}(d) provides a zoom in the low temperature region. We observe that below $T \approx 0.43$ the viscosity from Green-Kubo is governed by stable modes. Interestingly, this scale is compatible with the mode coupling temperature of this system, $T_{\text{MC}} \approx 0.44$ confirming that this dynamical crossover in $\eta$ coincides with the topological transition in the potential energy landscape between minima dominated and saddles dominated dynamics  as derived in \color{black} \cite{PhysRevLett.85.5356,PhysRevLett.85.5360}.

Finally, to further support the universality of the crossover observed in the Cu-Zr metallic liquid and the KA model, we examine a third system: a covalently bonded Fe$_{80}$P$_{20}$ metallic liquid. In Fig~\ref{fig5}(e), we present the viscosity as a function of temperature $T$, obtained numerically via the Green-Kubo formalism. For comparison, we also plot the theoretical prediction from Eq.~\eqref{eq01}, computed using only the stable modes and only the unstable localized modes (ULINMs). Once again, the theoretical model based on ULINMs closely approximates the numerically obtained viscosity. In ~\ref{fig5}(f), we zoom in on the low-temperature regime, near the mode-coupling temperature of the system, $T_{\text{MC}}\approx 920$ K (indicated by the vertical red arrow). As in the other two systems, we observe a distinct crossover: at $T>T_{\text{MC}}$, the viscosity is predominantly governed by ULINMs, whereas below $T_{\text{MC}}$, stable modes begin to dominate. This behavior reinforces our previous findings and provides strong evidence for the universality of the observed crossover.

\begin{figure}
    \centering
    \includegraphics[width=0.9\linewidth]{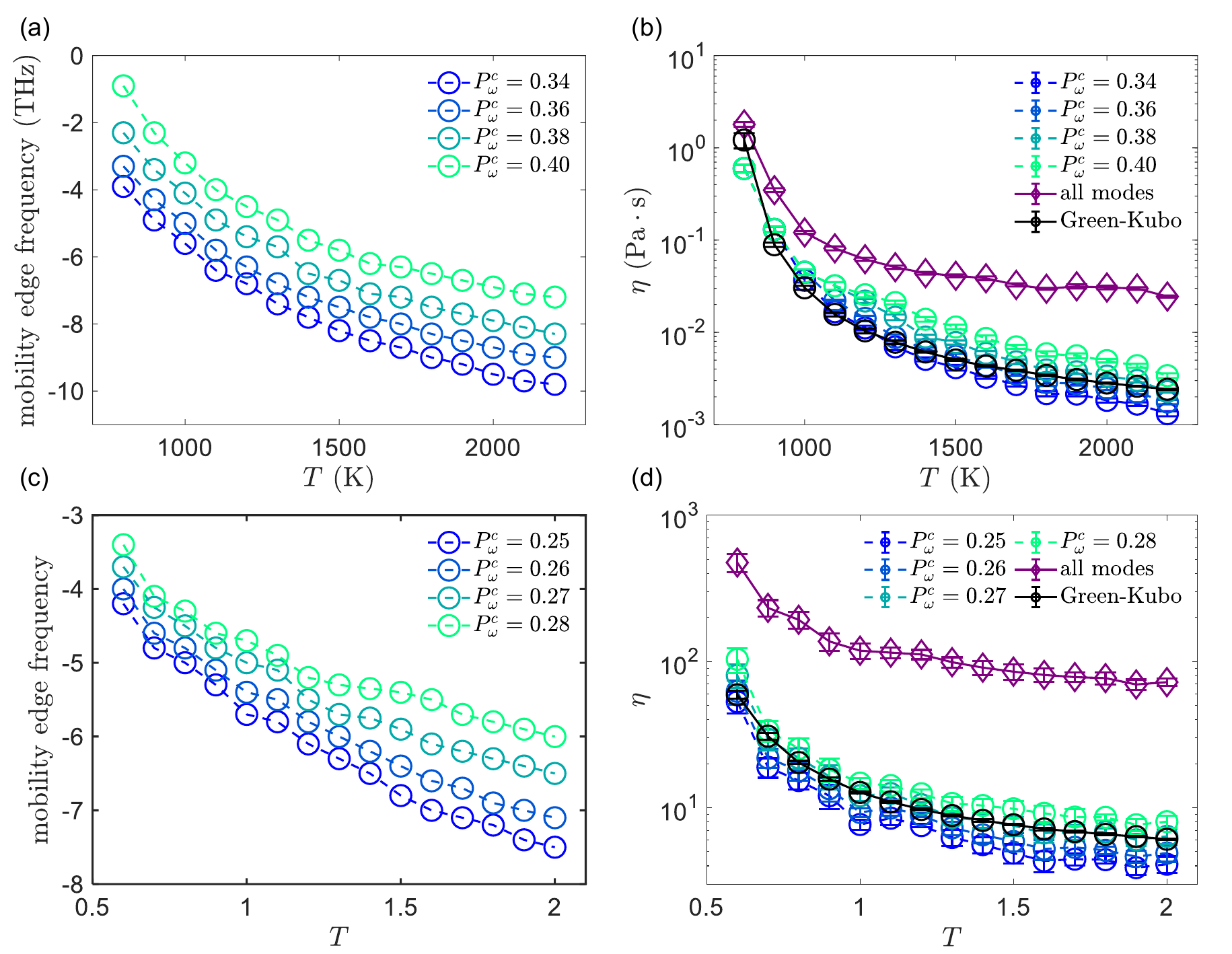}
    \caption{{\textbf{Effect of the mobility edge on viscosity.} (a) CuZr liquid and (c) Kob–Andersen liquid: mobility-edge frequency as a function of temperature for different choices of the critical participation ratio.
(b) CuZr liquid and (d) Kob–Andersen liquid: viscosity computed using the corresponding mobility-edge frequencies. The viscosity obtained from the Green–Kubo approach (i.e., including all modes) is shown for comparison.}}
    \label{fig6}
\end{figure}

To assess the sensitivity of our results to the precise identification of the mobility edge, we systematically vary the critical participation ratio used to distinguish localized from delocalized modes. Specifically, in Fig.~\ref{fig6}(a) and (c), we consider $P_\omega^c = 0.34, 0.36, 0.38,$ and $0.40$ for the CuZr liquid, and $P_\omega^c = 0.25, 0.26, 0.27,$ and $0.28$ for the Kob--Andersen liquid, respectively, and extract the corresponding mobility-edge frequency as a function of temperature. As expected, decreasing the critical participation ratio shifts the mobility edge to lower frequencies, thereby modifying the frequency range over which localized modes contribute to the viscosity.

Using the resulting mobility-edge frequencies, we compute the viscosity arising from (i) unstable localized modes only and (ii) all vibrational modes, and compare both estimates with the Green--Kubo benchmark. As shown in Fig.~\ref{fig6}(b) and (d), the viscosity obtained from the unstable localized modes fluctuates around the Green--Kubo value. In contrast, the viscosity calculated by including all modes exceeds the Green--Kubo result by roughly an order of magnitude. These results demonstrate that, although the choice of critical participation ratio affects the mobility-edge frequency and leads to modest quantitative variations in the viscosity, the qualitative conclusion remains robust: unstable localized modes dominate the viscous response, while contributions from all modes strongly overestimate the viscosity.

\begin{figure}
\centering
\includegraphics[width=0.75\textwidth]{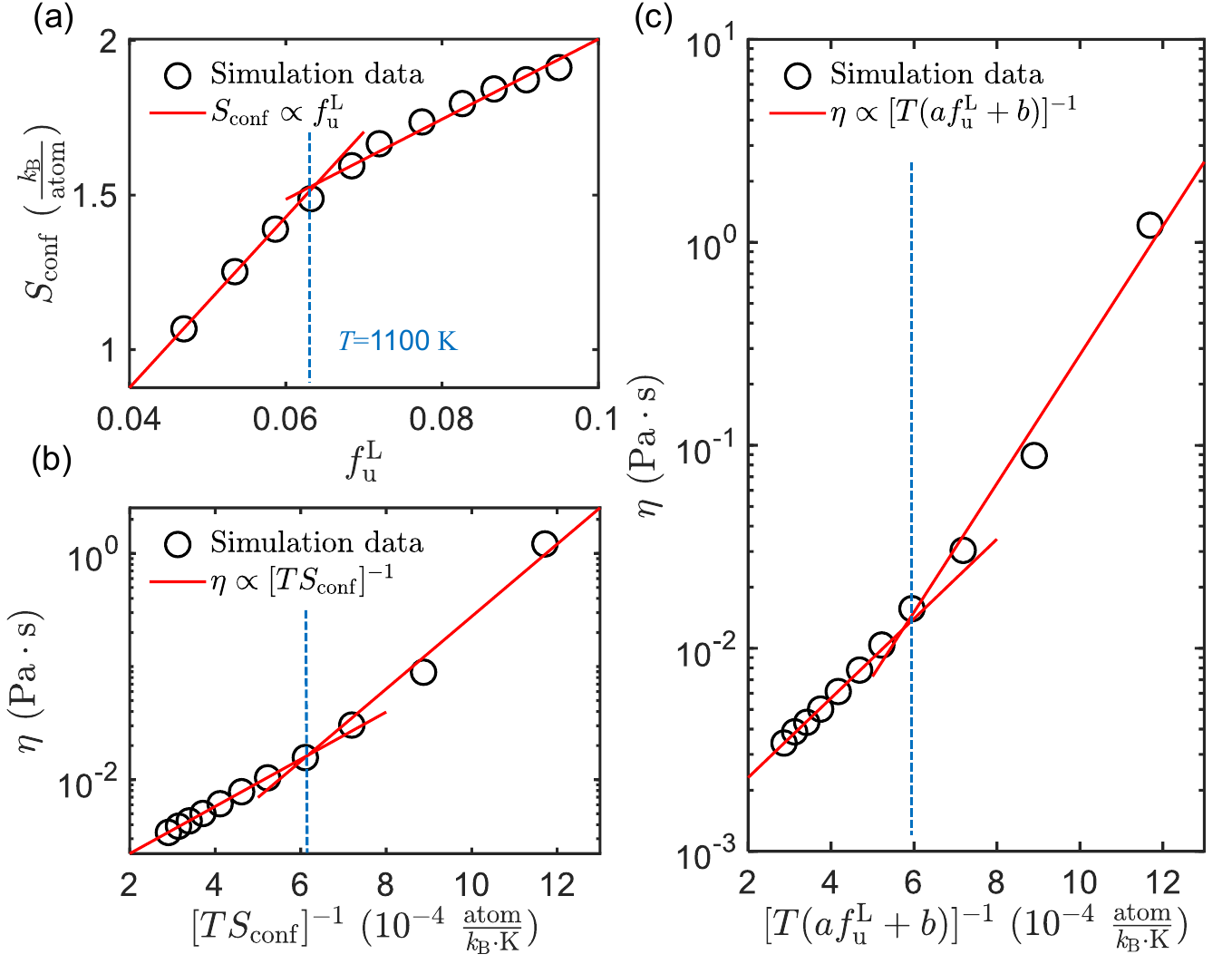}
\caption{\label{fig7}\textbf{Model connecting configurational entropy, instantaneous normal modes, and liquid viscosity.}
(a) Linear scaling between $S_{\mathrm{conf}}$ and $f_{\mathrm{u}}^{\mathrm{L}}$.
(b) Liquid viscosity as a function of $[T S_{\mathrm{conf}}]^{-1}$.
(c) Liquid viscosity as a function of $[T(a f_{\mathrm{u}}^{\mathrm{L}} + b)]^{-1}$.
In all panels, the solid lines represent model predictions assuming a crossover temperature of approximately $T = 1100$ K. This temperature coincides with the qualitative change in the localization nature of the INMs identified in Fig.~\ref{figS}.}
\end{figure}

Finally, inspired by the Adam--Gibbs theory~\cite{adam1965temperature} and motivated by recent proposals linking liquid diffusion to excess entropy~\cite{Douglass2024}, we construct a theoretical framework to quantitatively relate unstable localized instantaneous normal modes (ULINMs) to macroscopic viscosity. Following the spirit of Ref.~\cite{10.1063/1.3701564}, we attempt to connect ULINMs to the configurational entropy and, subsequently, to the viscous response.

The viscosity–INM correlation model is illustrated in Fig.~\ref{fig7}. As shown in Fig.~\ref{fig7}(a), the configurational entropy $S_{\mathrm{conf}}$ exhibits a two-stage linear dependence on $f_{\mathrm{u}}^{\mathrm{L}}$ over the temperature ranges of $800$–$1100$~K and $1200$–$1800$~K, respectively. Here, the configurational entropy is estimated by subtracting the vibrational entropy from the total entropy obtained via thermodynamic integration, following the procedure described in Ref.~\cite{doi:10.1021/acs.jpclett.3c03530}. The existence of two distinct scaling regimes indicates a change in the relationship between the configurational landscape and the population of unstable instantaneous normal modes. Notably, the intersection of these two regimes occurs at approximately $1100$~K.

\color{black}Furthermore, Fig.~\ref{fig7}(b) shows that the viscosity follows an Adam--Gibbs--like relation with the configurational entropy,
\begin{equation}
\eta \propto A \exp\!\left(\frac{B}{T S_{\mathrm{conf}}}\right),
\end{equation}
\color{black}which also displays a clear two-stage behavior. This observation mirrors the well-known connection between the structural $\alpha$-relaxation time and $S_{\mathrm{conf}}$, commonly interpreted within the random first-order transition framework~\cite{adam1965temperature,bouchaud2004adam,ozawa2019does}.

\color{black}By combining these two empirical relations, we arrive at
\begin{equation}
\eta \propto A \exp\!\left[\frac{B}{T\left(a f_{\mathrm{u}}^{\mathrm{L}}+b\right)}\right],
\end{equation}
\color{black}as validated by the simulation data shown in Fig.~\ref{fig7}(c). This result establishes a direct quantitative link between viscosity and ULINMs, reinforcing their central role in viscous transport. More broadly, it provides an intriguing bridge between microscopic vibrational features and macroscopic thermodynamics and dynamics in supercooled liquids.

In Fig.~7, we identify a crossover at $T \approx 1.6T_{\mathrm{g}}$ in the viscosity--entropy--mode correlation. To the best of our knowledge, this temperature does not coincide with any previously established characteristic scale. Specifically, it is distinct from both the mode-coupling temperature $T_{\mathrm{MC}} \approx 1.2T_{\mathrm{g}}$ \cite{PhysRevE.67.031507} and the dynamic onset temperature $T_{\mathrm{A}} \approx 2T_{\mathrm{g}}$ \cite{blodgett2015proposal,10.1063/1.4952986,REN2021113926}. Unlike $T_{\mathrm{A}}$, which marks the onset of nontrivial dynamics and the first deviation from simple liquid behavior, the temperature $T \approx 1.6T_{\mathrm{g}}$ corresponds to a qualitative change in the nature of the INMs. As shown in Fig.~\ref{fig02}(d), the fraction of delocalized unstable modes, $f_{\mathrm{u}}^{\mathrm{D}}$, decreases significantly with decreasing temperature, whereas the fraction of localized unstable modes, $f_{\mathrm{u}}^{\mathrm{L}}$, exhibits only a weak temperature dependence. This indicates that unstable modes evolve from extended directions characteristic of a homogeneous liquid to increasingly localized excitations. To further support this interpretation, we plot in Fig.~\ref{figS} the ratio of localized to extended unstable modes. As the temperature decreases, localized unstable modes become progressively dominant, with an inflection point observed near $1135~\mathrm{K}$, identified from the crossover between empirical fits in the low- and high-temperature regimes. This crossover might suggest that the spatial character of unstable modes is closely linked to the development of dynamic heterogeneity in the supercooled liquid, warranting further investigation.

\color{black}\section{Concluding remarks}

In this work, we have performed MD simulations combined with instantaneous normal mode analysis to investigate the microscopic origin of viscosity in liquids in terms of atomic motion, and its decomposition in terms of normal modes. Our results indicate that, aside from very close to the glass transition, and in fact below the mode-coupling temperature, unstable localized INMs are the microscopic excitations responsible for this process. 

Aside from providing a spectral decomposition of liquid viscosity into INMs, our results contribute to answering the question of what \textit{unstable INMs are and are not}. Unstable delocalized modes are the building blocks for liquid self-diffusion, while the complementary ULINMs appear as the microscopic facilitators underlying liquid viscosity at high-enough temperatures. 

Interestingly, this answer complements several previous analyses. In particular, in Ref.~\cite{Widmer-Cooper2008} it was shown that localized soft modes are the origin of irreversible structural reorganization. Here, we give a microscopic definition of these localized soft modes and we propose that they should be identified with ULINMs. This is also consistent with the results of \cite{10.1063/1.5127821} that showed that ULINMs act as facilitators of the liquid dynamics, located at the boundary regions between mobile and immobile clusters generated by dynamical heterogeneity. Moreover, our findings are aligned with the idea that, in glass-forming liquid, excitations are localized and relaxation is hierarchical \cite{PhysRevX.1.021013}. Finally, viscosity being governed by localized modes is also reflected in the recent results of \cite{doi:10.1073/pnas.2400611121}, where it was found that the change in the local barrier energy caused by the fundamental rearrangement or excitation of a \textit{few} particles (hence, localized modes) determines the change in the activation energy and the dynamics of the supercooled liquids. 

In summary, our analysis provides a spectral decomposition of liquid viscosity into individual instantaneous normal modes that capture the underlying microscopic atomic motion. This decomposition not only identifies the microscopic carriers of momentum transport in liquids but also opens a new path toward a quantitative, microscopic theory of liquid viscosity, thereby complementing earlier successful approaches~\cite{viswanath2007viscosity}.

\printcredits

\section*{Declaration of competing interest}
The authors declare that they have no known competing financial interests or personal relationships that could have appeared to influence the work reported in this paper.

\section*{Data availability}
Data will be made available on request.

\section*{Acknowledgments}
We thank A.~Zaccone and V.~Vaibhav for collaboration at an early stage of this work. We would like to thank J.~Douglas, M.~Pica Ciamarra, J.~Dyre, T.~Keyes, J.~Moon and A.~Lemaitre for useful comments and discussion.
This work was financially supported by the National Key R\&D Program of China (grant nos. 2025ZD0122000 and 2024YFA1208003), the Strategic Priority Research Program (grant nos. XDB0620103 and XDB0510301) of CAS, the National Outstanding Youth Science Fund Project of China (grant no. 12125206), and the National Natural Science Foundation of China (grant no. 12472112).
B.C. also acknowledges the support of the start-up funding from the Chinese University of Hong Kong, Shenzhen (No. UDF01003468) and the Shenzhen city “Pengcheng Peacock” Talent Program.
M.B. acknowledges the support of the Shanghai Municipal Science and Technology Major Project (Grant No.2019SHZDZX01) and the sponsorship from the Yangyang Development Fund.

\section*{Appendix}
\appendix

\section{Instantaneous normal modes}
\begin{figure}[t!]
    \centering
    \includegraphics[width=.6\linewidth]{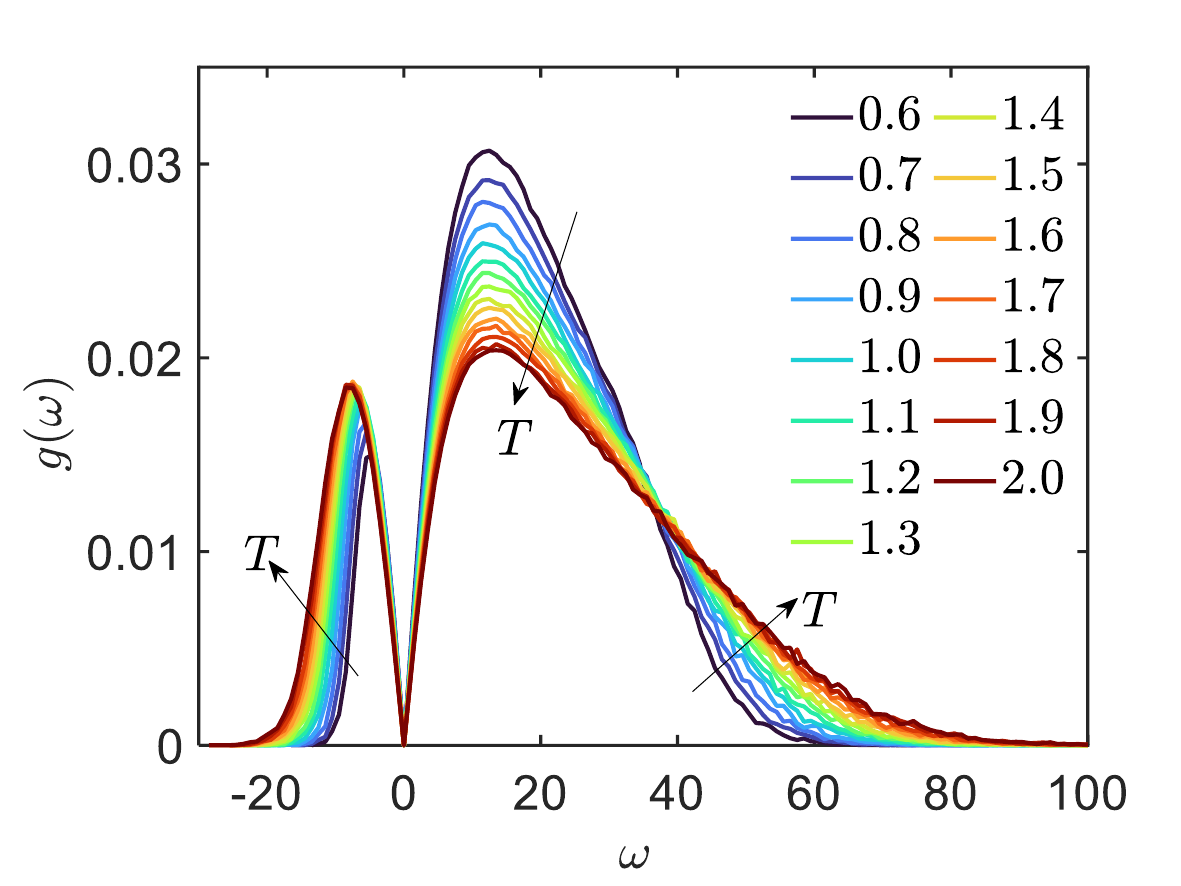}
    \caption{Instantaneous normal mode DOS for the Kob-Andersen model at different temperature. The arrows indicate the trend of change with increasing temperature.}
    \label{fig8}
\end{figure}

The dynamical matrix for the instantaneous liquid configuration at a given temperature (instantaneous Hessian matrix) is given by
\begin{equation}
\mathbf{D}=\frac{\partial^{2} U}{\partial r_{i, \alpha} \partial r_{j, \beta} \sqrt{m_{i} m_{j}}}
\label{eqs02},
\end{equation}
where $m_{i}$ is the mass of particle $i$ and $r_{i,\alpha}$ is the coordinate ($x ,y$ or $z$) of particle $i$. We directly diagonalize the dynamical matrix and calculate the vibrational density of states as
\begin{equation}
g(\omega)=\lim_{\Delta\omega\to0}\frac{\Delta n}{\Delta\omega}=\frac{1}{3N-3}\sum_{p}\delta(\omega-\omega_{p})
\label{eqs03},
\end{equation}
where $\omega_{p}$ is the eigenfrequency. The results for the INM DOS are shown in Fig. \ref{fig8} and present similar features as for the metallic liquid discussed in the main text.

\begin{figure}
    \centering
    \includegraphics[width=0.6\linewidth]{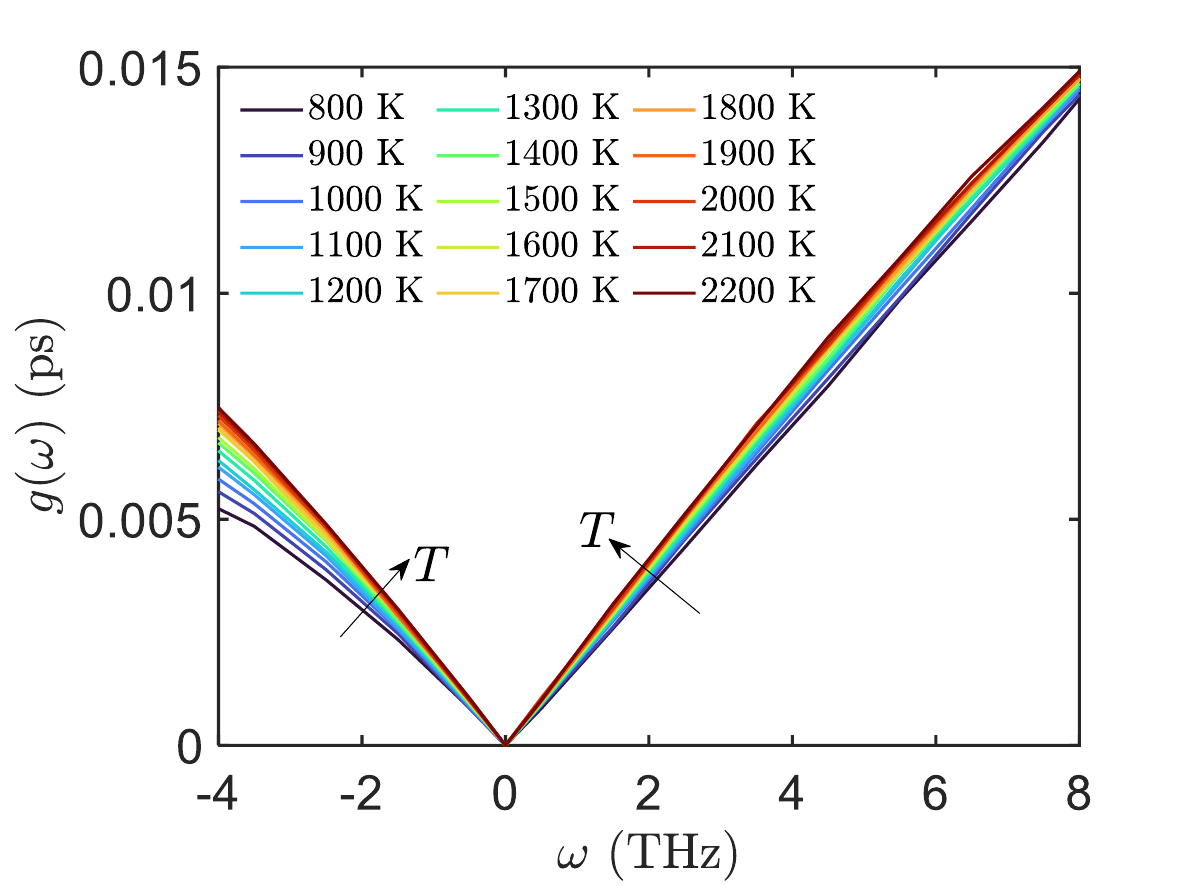}
    \caption{Zoom of the low frequency region of $g(\omega)$ for the $\mathrm{CuZr}$ metallic glass-forming liquid. A linear behavior $g(\omega)=a(T) |\omega|$ emerges at low frequency. From our simulation data, it is also evident that $a(T)$ grows with $T$.}
    \label{fig9}
\end{figure}

A zoom of the low frequency region of $g(\omega)$ for the  $\mathrm{CuZr}$ metallic glass-forming liquid is shown in Fig. \ref{fig9}. The data confirm the linear scaling $g(\omega)=a(T) |\omega|$ (with $a(T)$ growing with temperature) discussed in the main text.

\section{Affine force field correlator}\label{afaf}
\begin{figure}
    \centering
    \includegraphics[width=.6\linewidth]{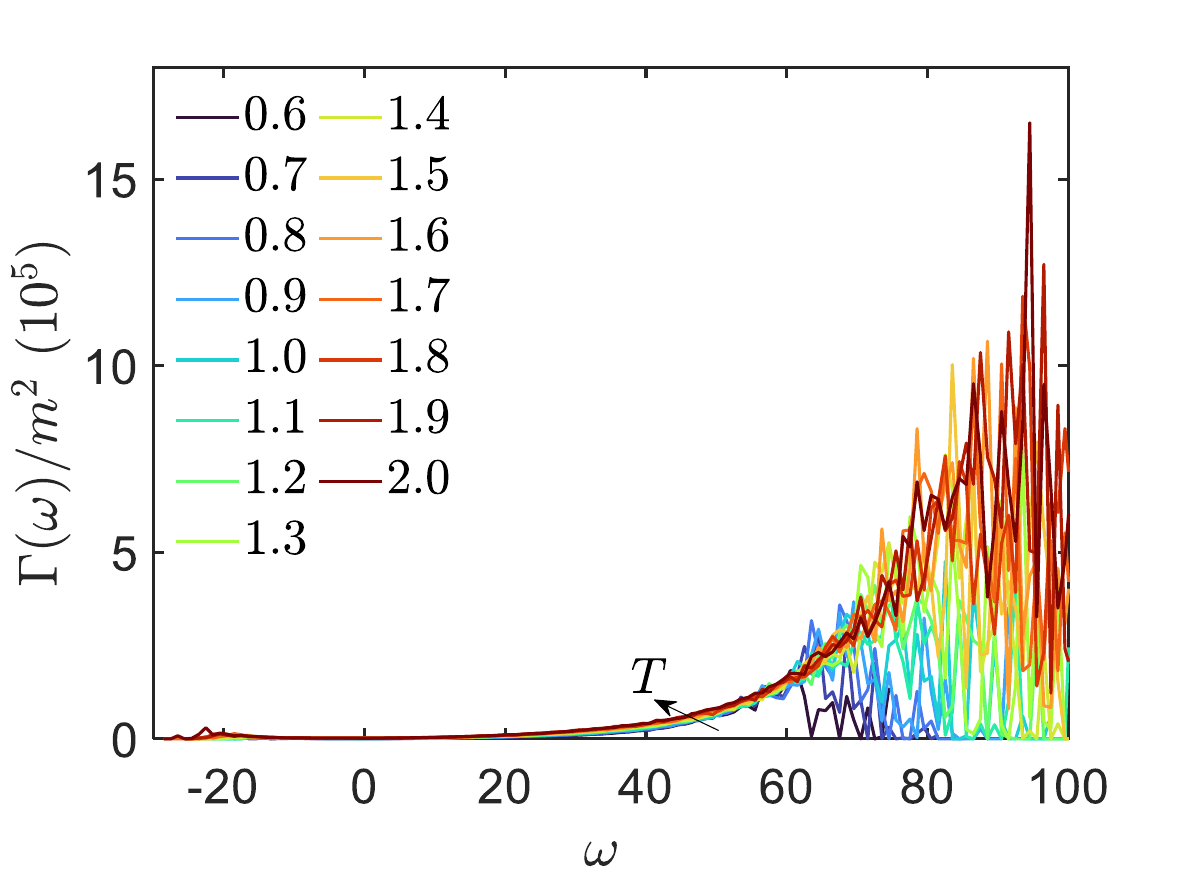}
    \caption{Affine force field correlator versus frequency at different temperature for the Kob-Andersen model. All the curves are obtained by averaging over 10 independent samples.}
    \label{fig10}
\end{figure}

According to nonaffine lattice dynamics theory \cite{lemaitre2006sum}, the affine force field is obtained from
\begin{equation}
\mathbf{\xi}_{i,xy}=\frac{\partial\mathbf{f}_{i}}{\partial\varepsilon _{xy}}\Bigg|_{\varepsilon \to0}=\frac{\mathbf{f}_{i}^{1}-\mathbf{f}_{i}^{0}}{\varepsilon  _{xy}},
\end{equation}
where $\mathbf{f}_{i}^{1}$ is the force on the $i$ atom in deformed state; $\mathbf{f}_{i}^{0}$ is the force on the $i$ atom in undeformed state; $\varepsilon  _{xy}$ is the simple shear strain. The projection of the eigenvectors $\mathbf{v}^{\omega_{p}}$ of the Hessian matrix onto the affine force field are then given by
\begin{equation}
\hat{\Xi}_{p,xy}=\mathbf{v}^{\omega_{p}}\cdot\mathbf{\xi}_{\mathrm{xy}},
\end{equation}
and the affine force field correlator is derived from
\begin{equation}
\Gamma(\omega)=\left\langle\hat{\Xi}_{p,xy}^{2}\right\rangle_{\omega_{p}\epsilon[\omega,\omega+d\omega]}. 
\end{equation}
A shear strain of 0.0001 is applied in our simulation. The affine force field correlator of the Kob-Andersen model is shown as a function of frequency, for the liquid sample at different temperature as shown in the Fig. \ref{fig10}.

\begin{figure}
    \centering
    \includegraphics[width=1\linewidth]{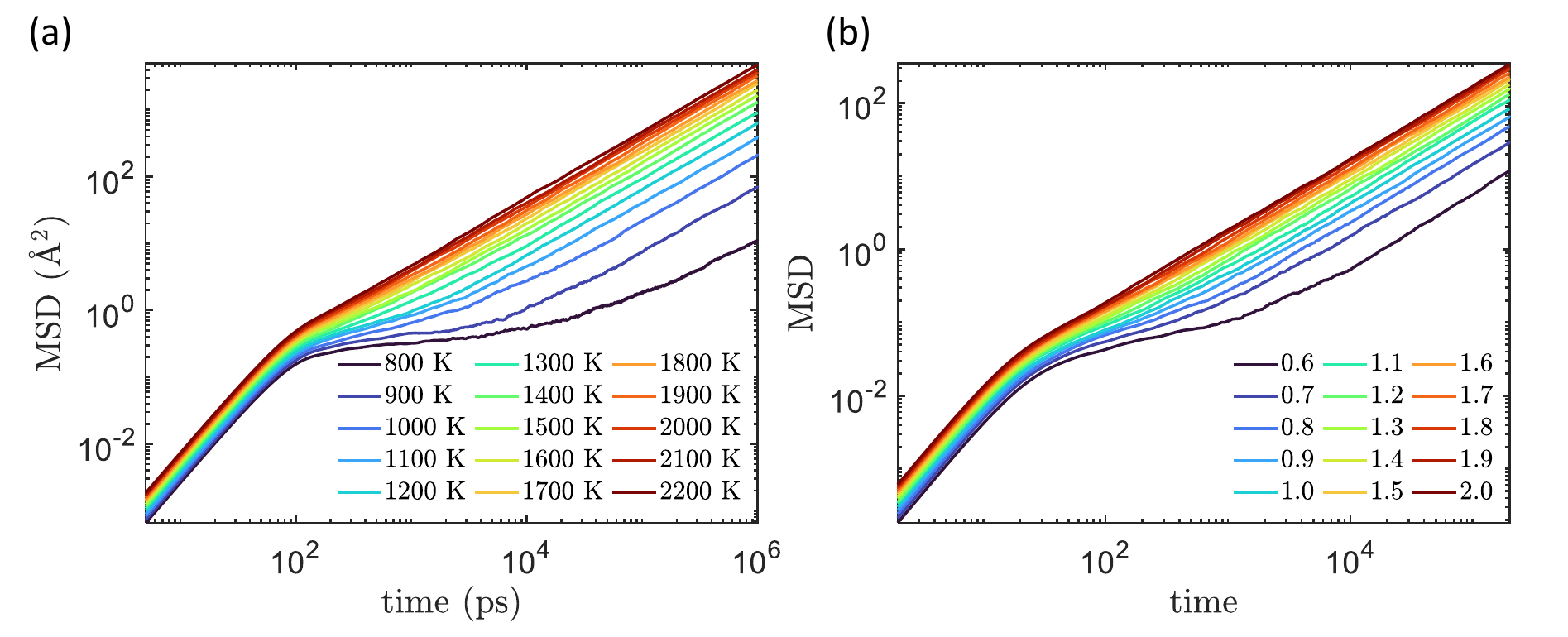}
    \caption{The mean square displacement as a function of time. \textbf{(a)} $\mathrm{CuZr}$ metallic liquid and \textbf{(b)} Kob-Andersen model. }
    \label{fig11}
\end{figure}

\section{Mean square displacement}

The diffusion constant can be obtained from the mean square displacement (MSD) as a function of time
\begin{equation}
D_{\alpha}=\frac{1}{2} \lim _{\tau \rightarrow \infty} \frac{\left\langle\frac{1}{N}\sum_{i}^{N}  \left[r_{i, \alpha}(t+\tau)-r_{i, \alpha}(t)\right]^{2}\right\rangle}{\tau}
\label{eqs04}.
\end{equation}
The average diffusion constant in an isotropic system in 3D is given by
\begin{equation}
D=\frac{1}{3} \sum_{\alpha =1}^{3} D_{\alpha } 
\label{eqs004}.
\end{equation}
The MSD of the $\mathrm{CuZr}$ system and the KA model at different temperatures are shown in Fig. \ref{fig11}.

\begin{figure}
    \centering
    \includegraphics[width=.6\linewidth]{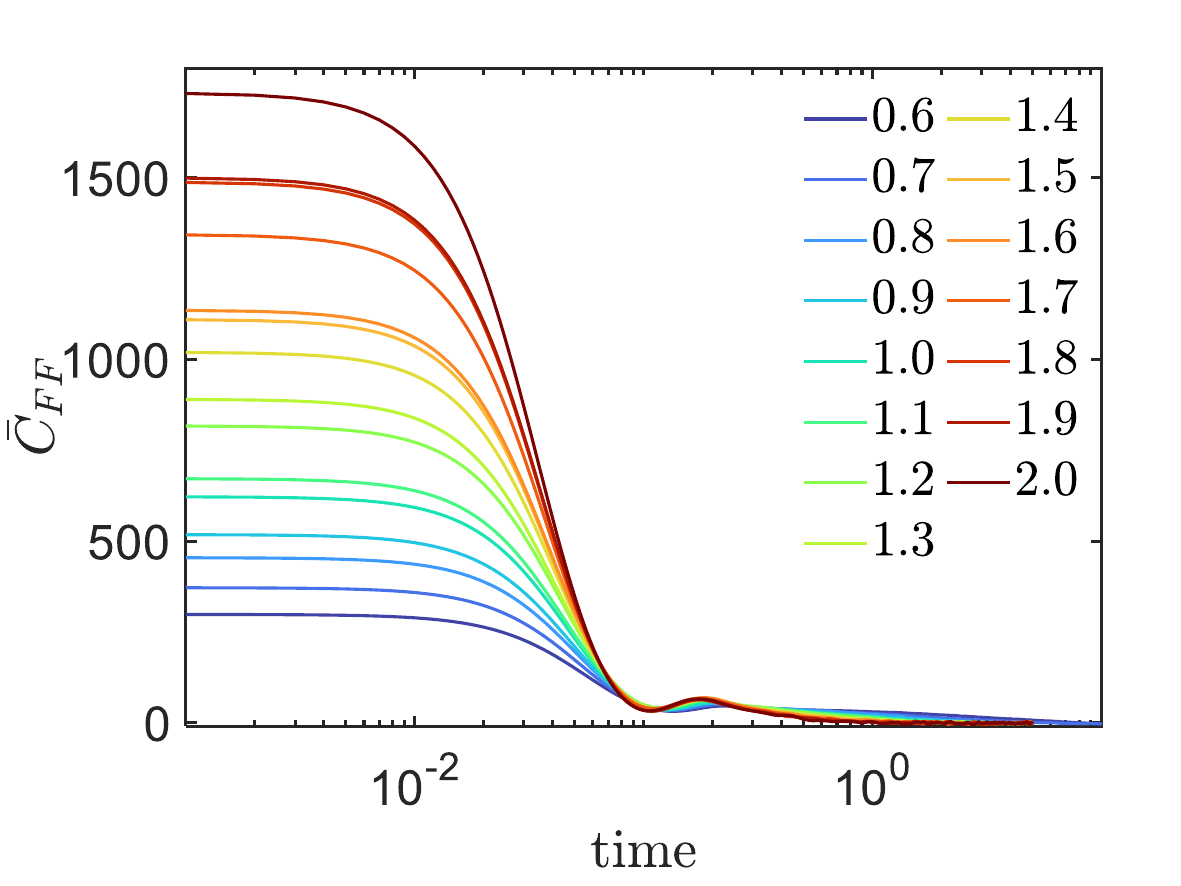}
    \caption{Auto-correlation function versus time at different temperatures $T \in [0.6,2]$ for the KA model. LJ units are used for the temperature $T$.}
    \label{fig12}
\end{figure}

\begin{figure}
    \centering
    \includegraphics[width=1\linewidth]{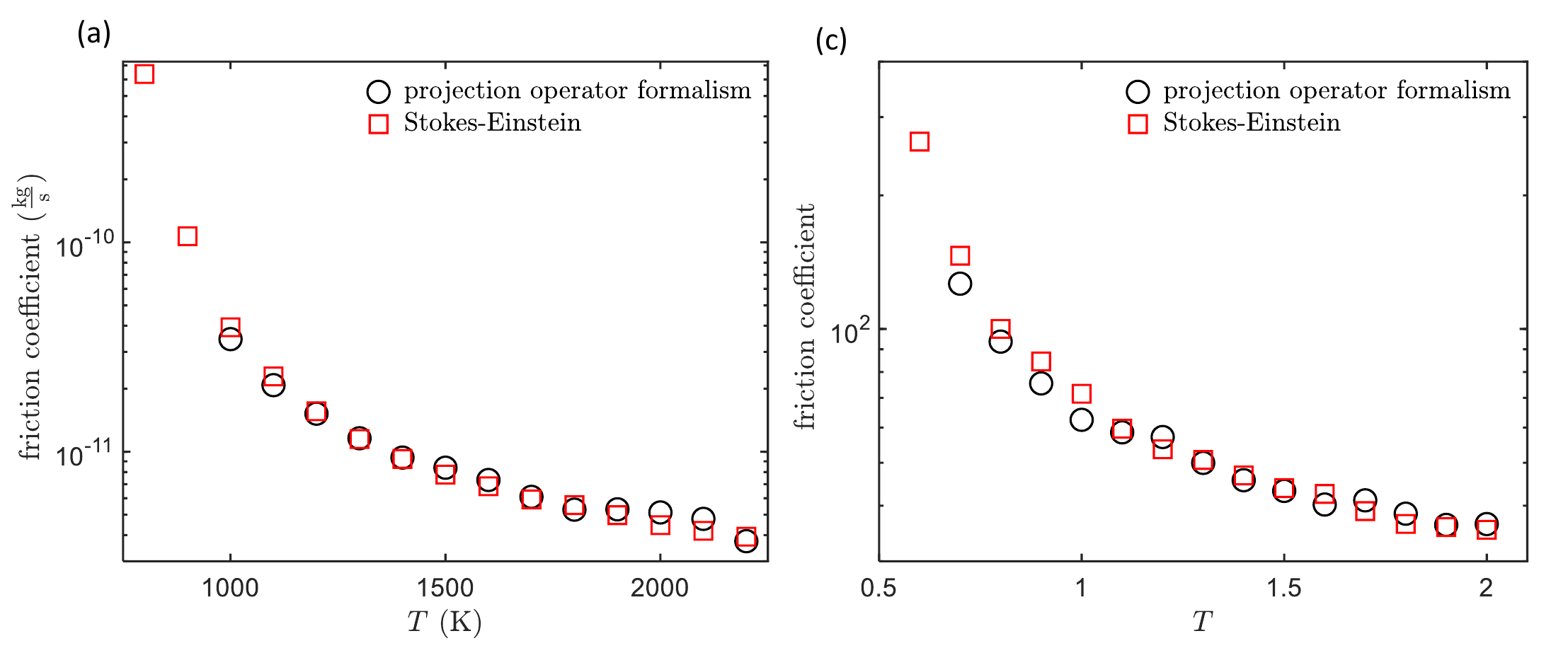}
    \caption{Consistency of $\int_{0}^{\infty}\bar{C}_{\mathbf{FF}}(t)/k_{\mathrm{B}}Tdt$ and $k_{\mathrm{B}}T/D$ for calculating the friction coefficient. \textbf{(a)} $\mathrm{CuZr}$ glass-forming liquid and \textbf{(b)} Kob-Andersen model.}
    \label{fig13}
\end{figure}

\section{Projection operator technique}\label{propro}

According to the projection operator technique \cite{10.1063/1.4868653,PhysRevLett.116.147804,Jung2017IterativeRO}, the memory kernel 
\begin{equation}
\upsilon(t)=\frac{\left\langle \mathbf{F}\left(e^{-i(1-\mathcal{P})\mathcal{L}t}\mathbf{F}\right)\right\rangle}{k_\text{B}T}=\frac{\langle \mathbf{FF}_r(t)\rangle}{k_\text{B}T}=\frac{\bar{C}_{\mathbf{FF}}(t)}{k_\text{B}T}
\label{eq0004},
\end{equation}
where $\mathbf{F}$ the true force acting on the tagged particle. Based on the second-order backward algorithm, $\bar{C}_{\mathbf{AA}}(t)=\left \langle \mathbf{A}_{0}\mathbf{A}_{t}^{-}  \right \rangle $, where $\mathbf{A}$ is an observable, and $\mathbf{A}_{t}^{-}=e^{-i(1-\mathcal{P})\mathcal{L}t}\mathbf{A}_{0}$. $\mathbf{F}_{t}^{-}$ evolves according to the backward orthogonal dynamics
\begin{equation}
\mathbf{F}_{t+\delta t}^{-}(\mathbf{q},\mathbf{p})=\mathbf{F}_{-t}^{-}\left(\mathbf{q}^{\delta t},\mathbf{p}^{\delta t}\right)+\int_{0}^{\delta t}\mathbf{F}_{-u}\left(\mathbf{q}^{\delta t},\mathbf{p}^{\delta t}\right)\frac{\left\langle\mathbf{P}_{0}\mathbf{F}_{t-u}^{-}\right\rangle}{\left\langle\mathbf{P}_{0}^{2}\right\rangle}du
\label{eqs05},
\end{equation}
where $\mathbf{q}$ and $\mathbf{p}$ are respectively the atomic positions and  momenta. The second-order propagator
\begin{equation}
F_{n+1}^{-}(m)=F_{n}^{-}(m)+\beta(n)F_{n}(m)\frac{\delta t}{2}+\frac{F_{n}(m)}{1-\frac{\delta t}{2}\kappa(n)}\biggl[\gamma(n)+\Delta(n)\beta(n)\frac{\delta t}{2}\biggr]\frac{\delta t}{2}
\label{eqs06},
\end{equation}
with
\begin{equation}
\beta(n)=\frac{\sum_{m=1}^{N_{\mathrm{traj}}-N_{\mathrm{corr}}}P_{n}(m)F_{n}^{-}(m)}{\sum_{m=1}^{N_{\mathrm{traj}}-N_{\mathrm{corr}}}P_{n}(m)^{2}}, \qquad 
\kappa(n)=\frac{\sum_{m=1}^{N_{\mathrm{traj}}-N_{\mathrm{corr}}}F_{n}(m)P_{n}(m)}{\sum_{m=1}^{N_{\mathrm{traj}}-N_{\mathrm{corr}}}P_{n}(m)^{2}},
\label{eqs08}
\end{equation}

\begin{equation}
\gamma(n)=\frac{\sum_{m=1}^{N_{\mathrm{traj}}-N_{\mathrm{corr}}}P_{n}(m)F_{n}^{-}(m+1)}{\sum_{m=1}^{N_{\mathrm{traj}}-N_{\mathrm{corr}}}P_{n}(m)^{2}}, \qquad \Delta(n)=\frac{\sum_{m=1}^{N_{\mathrm{traj}}-N_{\mathrm{corr}}}F_{n}(m)P_{n}(m+1)}{\sum_{m=1}^{N_{\mathrm{traj}}-N_{\mathrm{corr}}}P_{n}(m)^{2}}
\label{eqs10}.
\end{equation}

As a result, based on the trajectories of the atoms in each configuration we can get $\mathbf{F}_{t}^{-}$. Finally, $\bar{C}_{FF}$ can be obtained,
\begin{equation}
\bar{C}_{FF}(n\delta t)=\frac{1}{N_{\mathrm{traj}}-N_{\mathrm{corr}}}\sum_{m=1}^{N_{\mathrm{traj}}-N_{\mathrm{corr}}}F_{n}(m)F_{n}^{-}(m).
\label{eqs11}
\end{equation}

The results for auto-correlation function with different temperatures of KA model are shown in the Fig. \ref{fig12} and present similar features as for the metallic liquid discussed in the main text. As shown explicitly in the Fig. \ref{fig13}, the results for the friction coefficient obtained using $\int_{0}^{\infty}\bar{C}_{\mathbf{FF}}(t)/k_{\mathrm{B}}Tdt$ and $k_{\mathrm{B}}T/D$ are in good agreement. Therefore, we will use $k_{\mathrm{B}}T/D$, instead of $\int_{0}^{\infty}\bar{C}_{\mathbf{FF}}(t)/k_{\mathrm{B}}Tdt$, to calculate $\tilde{\upsilon } (0)$ at low temperatures.

\section{Participation ratio}\label{prpr}

\begin{figure}
    \centering
    \includegraphics[width=1\linewidth]{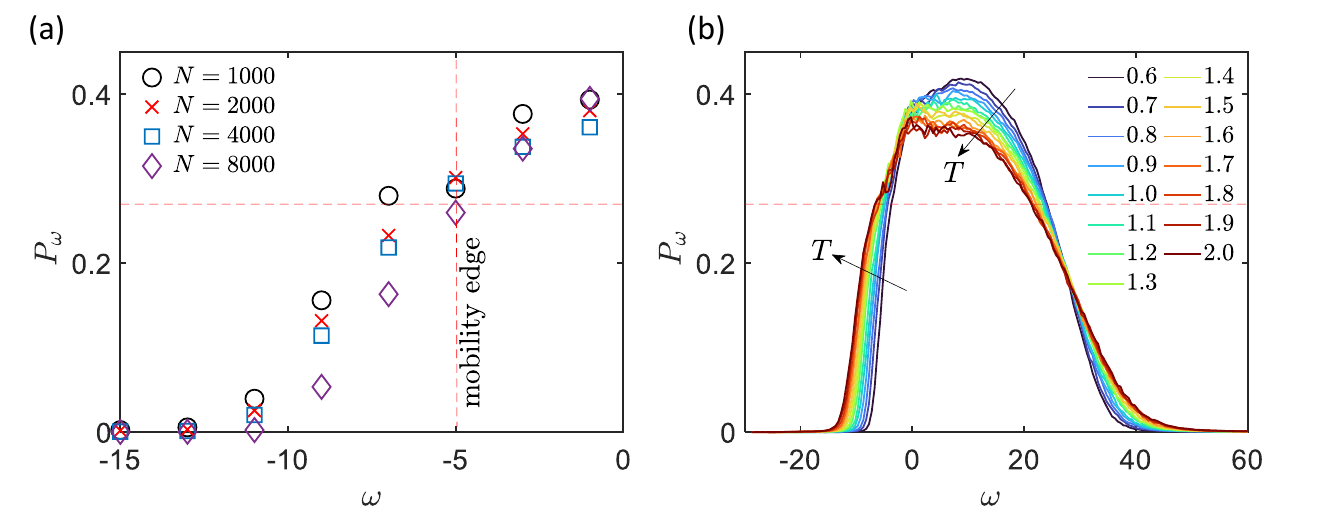}
    \caption{The participation ratio of Kob-Andersen model. \textbf{(a)} Variation of participation ratio with mode size at temperature 1; \textbf{(b)} Participation ratio versus frequency at different temperature. The dotted horizontal line represents $P_{\omega}=0.27$.}
    \label{fig14}
\end{figure}

The participation ratio of Kob-Andersen model is calculated and analyzed in Fig. \ref{fig14} and displays features similar to the case of the $\mathrm{CuZr}$ metallic glass-forming liquid discussed in the main text.

\begin{figure}
    \centering
    \includegraphics[width=0.6\linewidth]{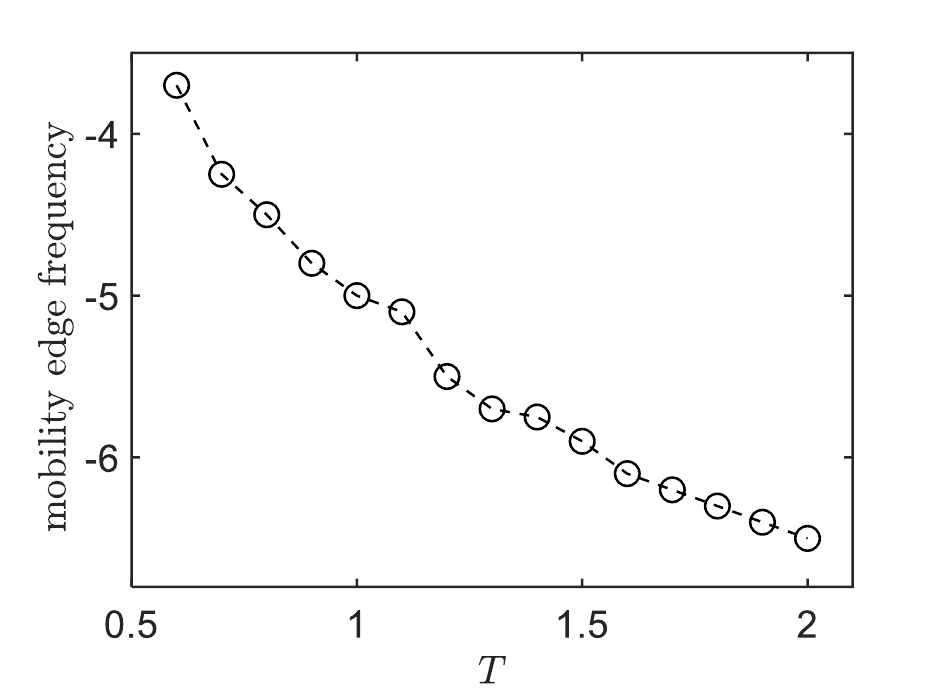}
    \caption{Temperature dependence of the mobility edge frequency for the Kob-Andersen model.}
    \label{fig15}
\end{figure}

 As done in the main text, we can calculate the dependence of the mobility edge frequency on temperature for the KA model, as shown in the Fig. \ref{fig15}.

\begin{figure}
    \centering
    \includegraphics[width=1\linewidth]{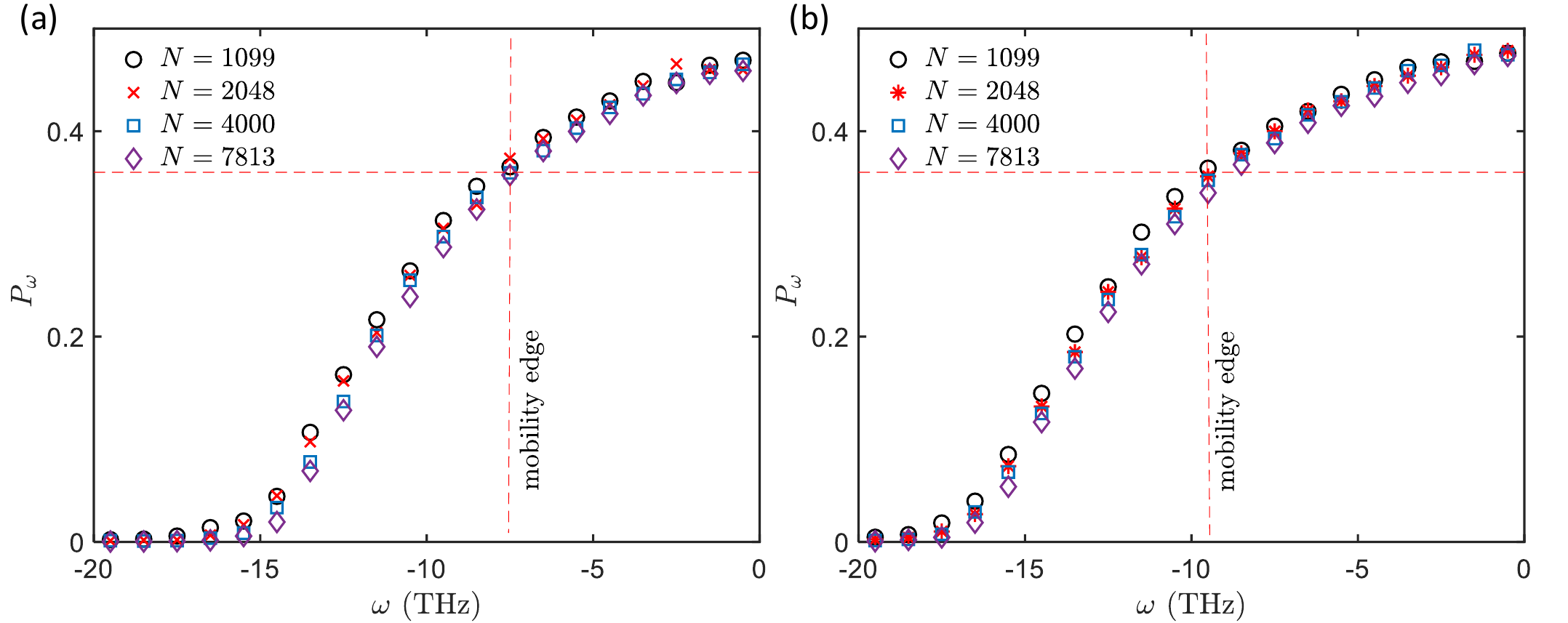}
    \caption{Variation of the participation ratio of $\mathrm{CuZr}$ metallic glass-forming liquid as a function of (imaginary frequency) at \textbf{(a)} $1500$ K and \textbf{(b)} $2100$ K upon changing the size of the simulation box.}
    \label{fig16}
\end{figure}

\begin{figure}
    \centering
    \includegraphics[width=0.8\linewidth]{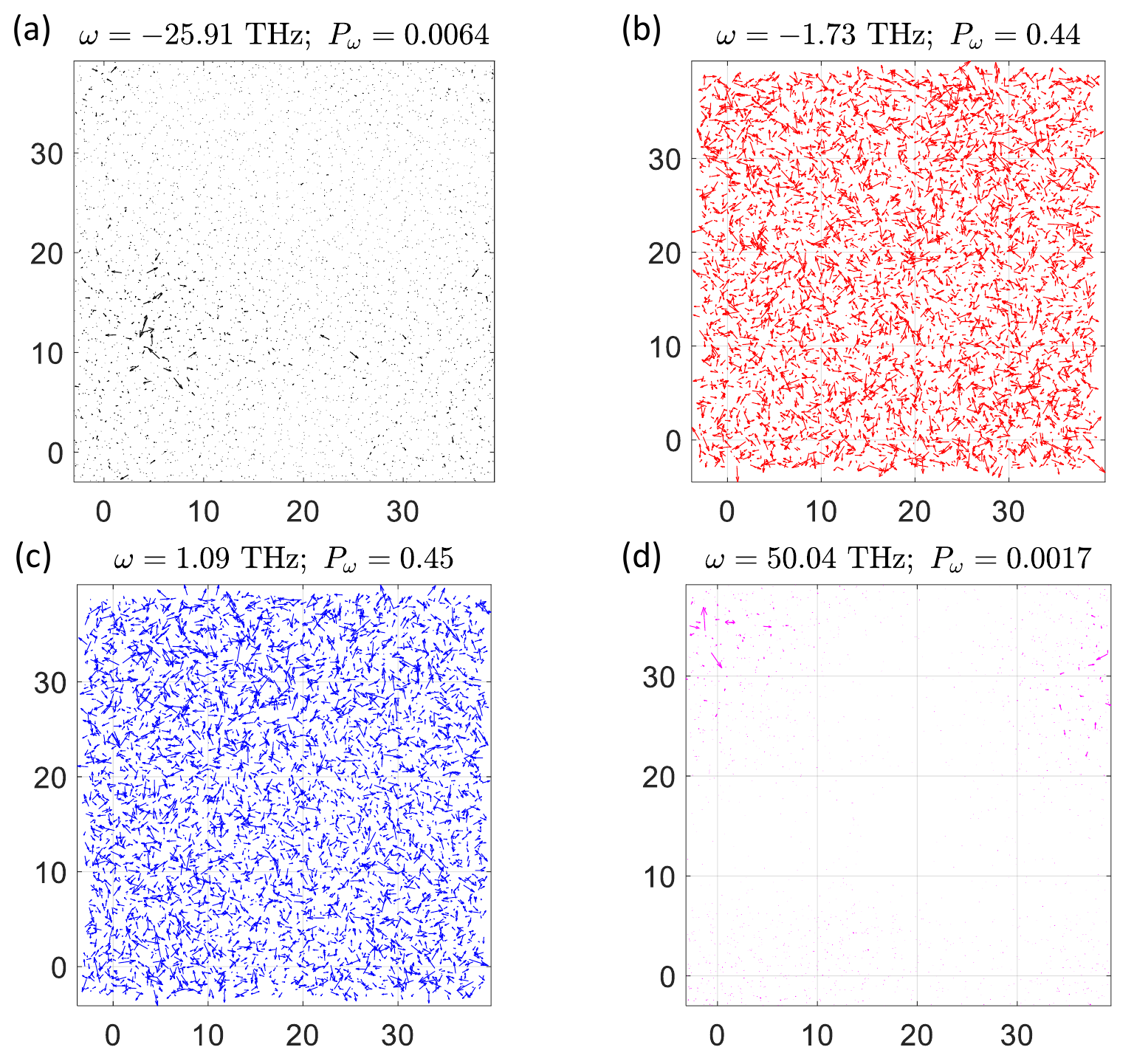}
    \caption{Spatial projection of the eigenvectors on the xy-plane of each particle corresponding to \textbf{(a)} unstable localized mode ($\omega=-25.91\ \mathrm{THz}$), \textbf{(b)} unstable delocalized mode ($\omega=-1.73\ \mathrm{THz}$), \textbf{(c)} stable delocalized mode ($\omega=1.09\ \mathrm{THz}$) and \textbf{(d)} stable localized mode ($\omega=50.04\ \mathrm{THz}$) of $\mathrm{CuZr}$ metallic liquid at 1500 K.}
    \label{fig17}
\end{figure}

Meanwhile, variation of the participation ratio of $\mathrm{CuZr}$ metallic glass-forming liquid as a function of (imaginary frequency) at $1500$ K and $2100$ K upon changing the size of the simulation box as shown in the Fig. \ref{fig16}. A typical localized and delocalized mode are shown by the magnitude of each particles vibrational vector as shown in the Fig. \ref{fig17}. Several high amplitude eigenvectors at local frequencies indicate strong localization, indicating that the energy associated with these modes is confined to a small spatial region.

\section{Bulk modulus}\label{bubu}

\begin{figure}[htbp]
    \centering
    \includegraphics[width=0.5\linewidth]{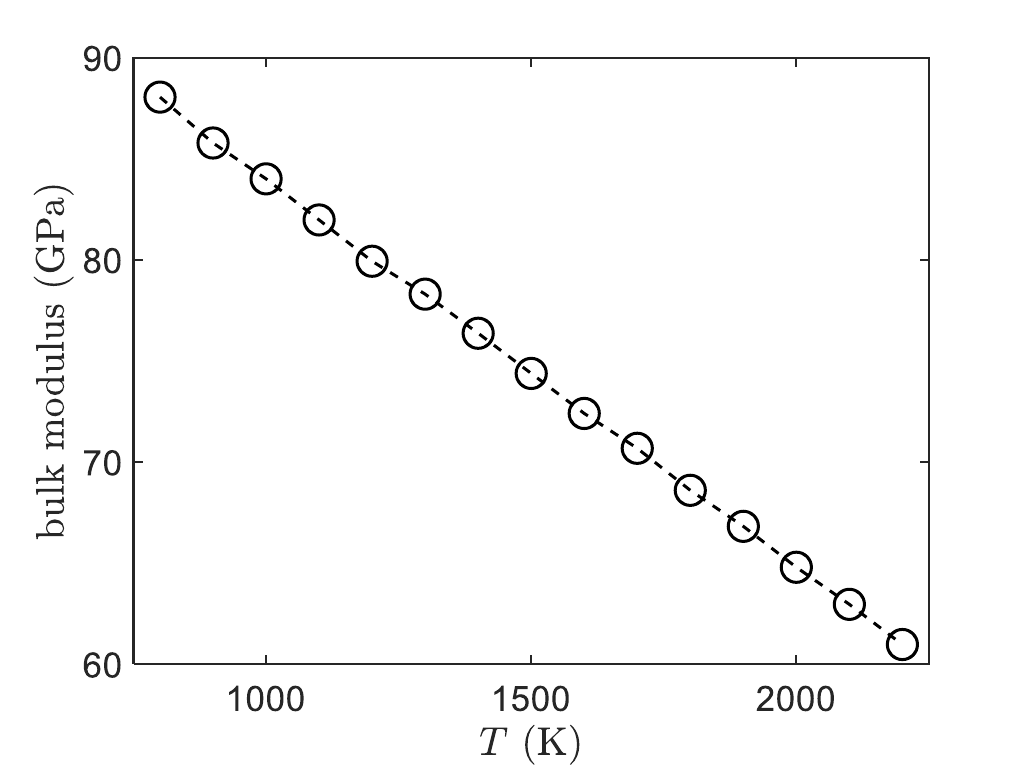}
    \caption{Temperature dependence of bulk modulus for the $\mathrm{CuZr}$ metallic glass-forming system.}
    \label{fig18}
\end{figure}

\begin{figure}[htbp]
    \centering
    \includegraphics[width=0.5\linewidth]{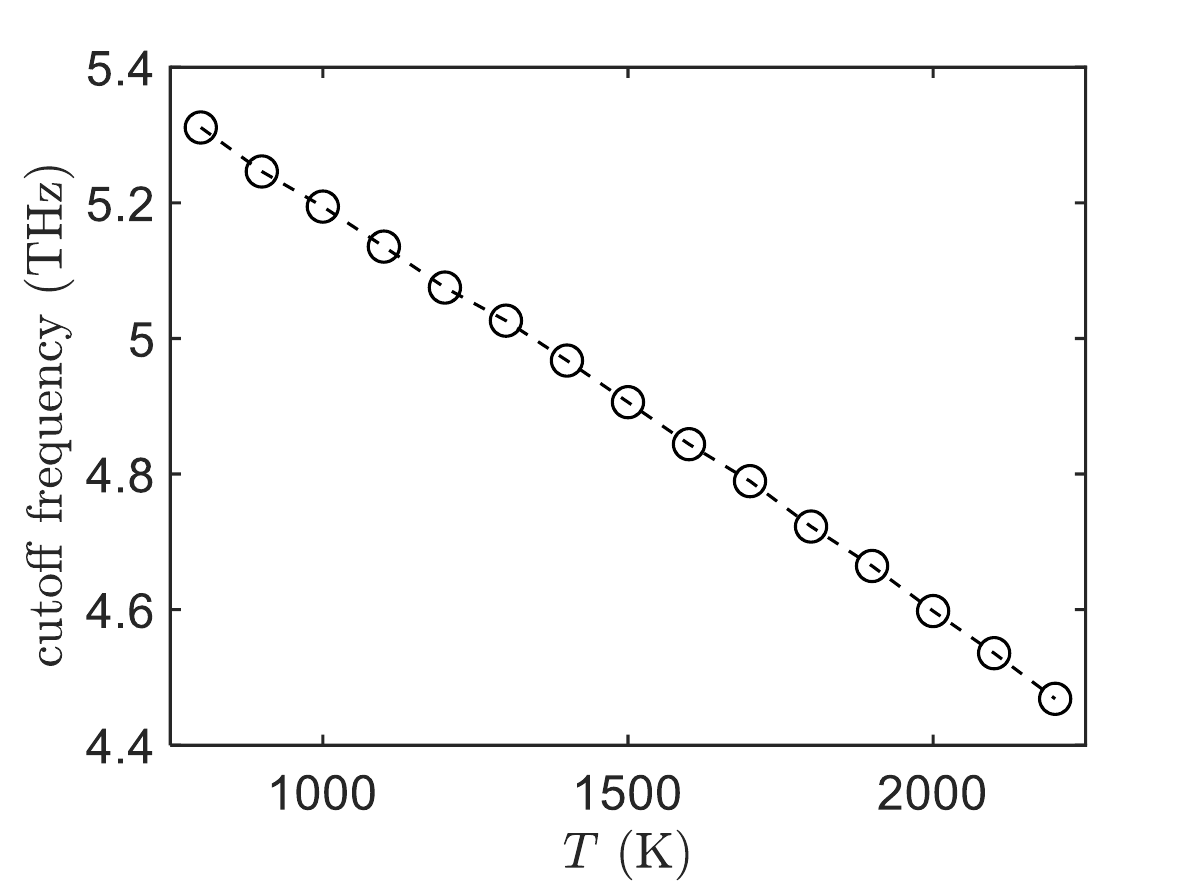}
    \caption{Temperature dependence of cutoff frequency $\omega_{\min }$ for the $\mathrm{CuZr}$ metallic glass-forming system.}
    \label{fig19}
\end{figure}

The elastic constants are calculated using LAMMPS \cite{lammps}, and then bulk modulus can be obtained according to the Voigt-Reuss-Hill approximation \cite{Hill_1952}. In the Fig. \ref{fig18}, it is shown that the bulk modulus of $\mathrm{CuZr}$ decreases with increasing temperature. Fig. \ref{fig19} shows the value of the cutoff frequency, $|\omega_{\min }|=(2 \pi \sqrt{B / \bar{\rho}}) / L$, as a function of temperature for the $\mathrm{CuZr}$ metallic liquid.

\section{Generalized Langevin equation (GLE)}

The linear generalized Langevin equation (GLE) is given by \cite{10.1063/1.4922941}
\begin{equation}
\dot{\mathbf{p}}(t)=-m \bar{\omega}^{2} \mathbf{q}(t)-\int_{0}^{t}\upsilon (t-\tau)\mathbf{v}(\tau) d \tau+\mathbf{F}_{r}(t)
\label{eqs18},
\end{equation}
where $\bar{\omega}$ is the frequency of an effective harmonic oscillator. Multiplying the linear GLE with $\mathbf{p}(0)$ and taking a canonical ensemble average, we can get
\begin{equation}
\dot{C}_{\mathbf{pp} }(t)=-m \bar{\omega}^{2} C_{\mathbf{pq} }(t)-\int_{0}^{t} \upsilon (t-\tau) C_{\mathbf{p v} }(\tau) d \tau
\label{eqs19}.
\end{equation}
By using the relation
\begin{equation}
C_{\mathbf{pq} }(t)=\int_{0}^{t} C_{\mathbf{p v} }(\tau) d \tau,
\label{eqs20}
\end{equation}
Eq. \ref{eqs19} can be simplified in the following form
\begin{equation}
\dot{C}_{\mathbf{pp} }(t)=-\int_{0}^{t} K (t-\tau) C_{\mathbf{p v} }(\tau) d \tau
\label{s21},
\end{equation}
where
\begin{equation}
K(t)=\upsilon (t)+m\bar{\omega}^{2}
\label{s22}.
\end{equation}
Upon Fourier transforming Eqs. \eqref{s21} and \eqref{s22}, one obtains
\begin{equation}
\tilde{K} (\omega )=\frac{C_{\mathbf{pp} }(0)}{\tilde{C}_{\mathbf{pv} }(\omega )}-i\omega
\label{s23},
\end{equation}
\begin{equation}
\tilde{K}(\omega)=\tilde{\upsilon}(\omega )+\pi m\bar{w}^2\delta(\omega)-i\frac{m\bar{w}^2}{\omega}
\label{s24}.
\end{equation}
Comparing Eqs. \eqref{s23} and \eqref{s24}, we finally obtain
\begin{equation}
\tilde{\upsilon}(0)=\lim_{\omega\to0}\tilde{\upsilon}(\omega )\approx\lim_{\omega\to0}\tilde{K}(\omega)=\lim_{\omega\to0}\frac{C_{{_{\mathbf{pp} }}}(0)}{\tilde{C}_{{_{\mathbf{pv} }}}(\omega)}=\frac{C_{{_{\mathbf{pp} }}}(0)}{\tilde{C}_{{_{\mathbf{pv} }}}(\omega)}=\frac{mk_{{_{\mathrm{B}}}}T}{m\tilde{C}_{{_{\mathbf{vv} }}}(0)}=\frac{k_{{_{\mathrm{B}}}}T}{\int_{0}^{\infty}\left \langle \mathbf{v(t)v(0)} \right \rangle dt}=\frac{k_{{_{\mathrm{B}}}}T}{D}
\label{s25}.
\end{equation}

\section{Verification of the liquid phase }
\begin{figure}
    \centering
    \includegraphics[width=0.5\linewidth]{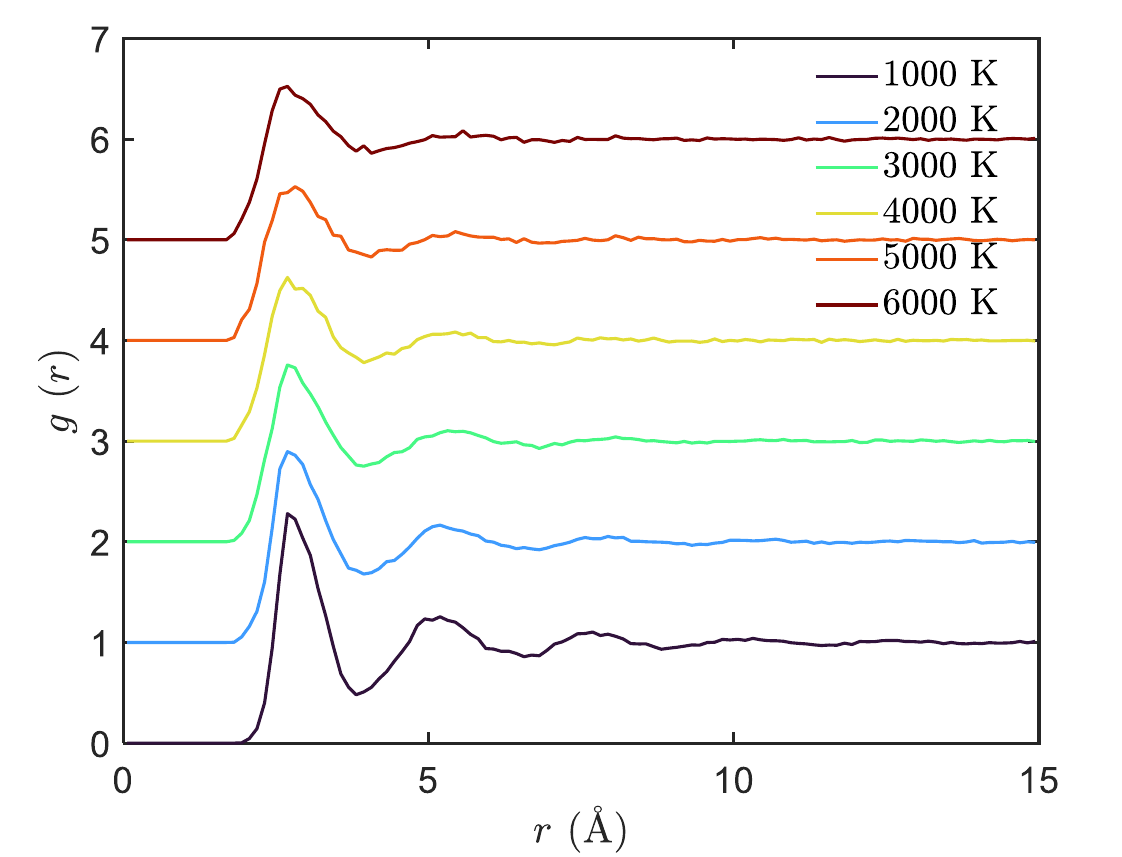}
    \vspace{0.2cm}
    \includegraphics[width=\linewidth]{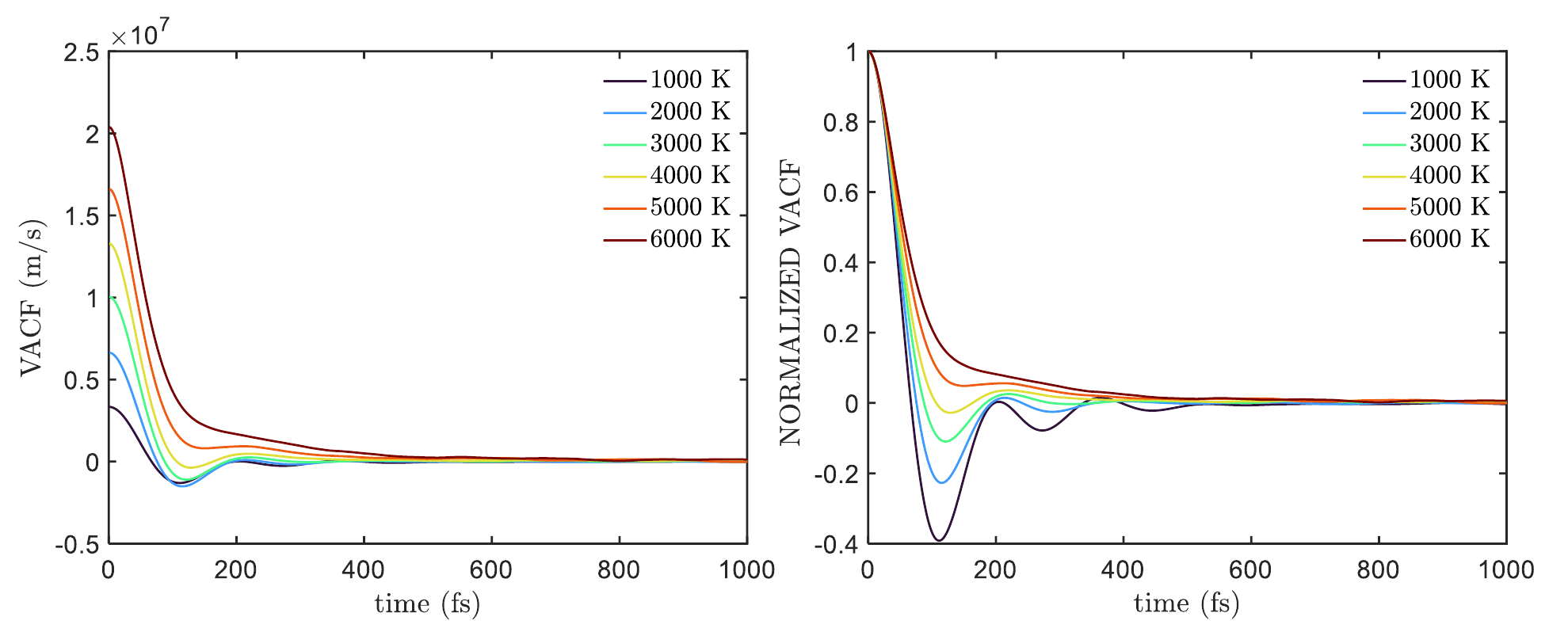}
    \caption{\textbf{Top panel:} pair distribution function $g(r)$ for the Cu-Zr metallic liquid at different temperatures. \textbf{Bottom panel:} velocity auto-correlation function VACF for the same system.}
    \label{fig20}
\end{figure}

Our computations for the Cu-Zr metallic liquid extend to high temperatures, $\approx 6000$ K. As evident from the results presented in the main text, the viscosity is perfectly captured by the ULINM even at high temperatures. Here, we show that, despite the large temperatures, the Cu-Zr metallic liquid is still in the liquid state. This is explicitly proven by both the pair distribution function $g(r)$ and the velocity autocorrelation function VACF shown in Fig. \ref{fig20}. As evident from there, below $6000$ K the system has still liquid characteristics. 

\section{Crossover temperature based on INMs}
\begin{figure}
    \centering
    \includegraphics[width=0.5\linewidth]{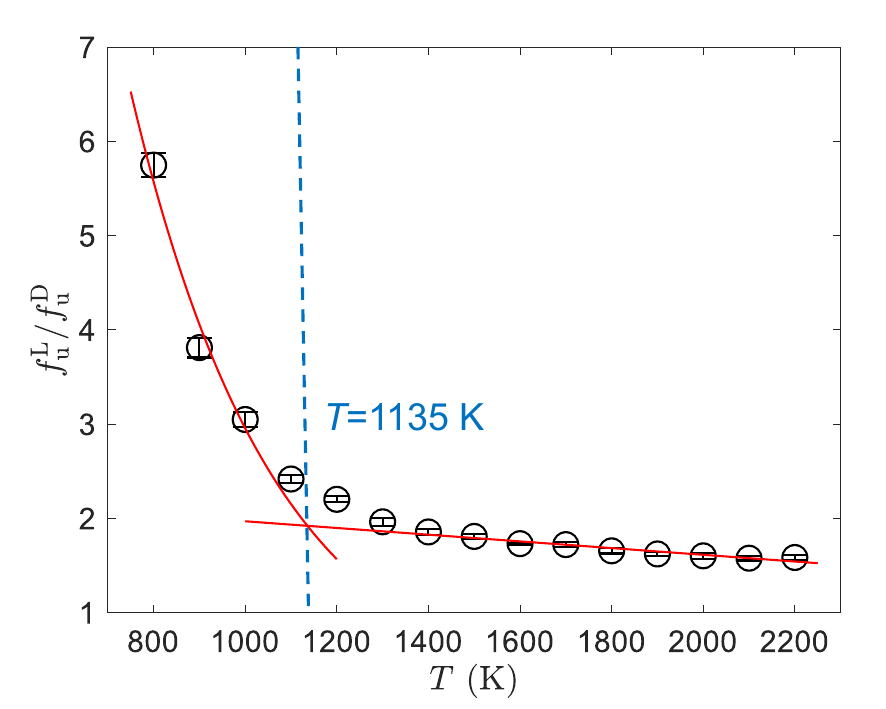}
    \caption{
    \textbf{Crossover temperature based on INMs.} Temperature dependence of the ratio of localized to extended unstable modes. A crossover is observed near $T \approx 1135\,\mathrm{K}$ (vertical dashed line), separating a low-temperature regime characterized by a rapid increase in localization from a high-temperature regime with weak temperature dependence. The low-temperature data are empirically fitted by an exponential function, while the high-temperature data are well described by a linear fit.}
    \label{figS}
\end{figure}

In the Fig. \ref{figS}, we investigate the ratio between localized unstable modes and extended modes. This analysis indicates that the identified crossover temperature is closely associated with a change in the spatial character of the INMs, specifically the increasing dominance of localized unstable modes.

\clearpage
\color{black}

\end{document}